\documentclass[aps,prd,preprint]{revtex4-2}
\usepackage{amsmath,amssymb,array,calc,rotating,epsfig,psfrag}
\usepackage{graphicx}
\usepackage{hyperref}
\usepackage{mathtools,slashed}
\usepackage[utf8]{inputenc} 
\usepackage[dvipsnames]{xcolor}
\usepackage{tensind}
\usepackage{cancel}
\usepackage{tensor}
\usepackage{cleveref}
\usepackage{multirow}
\tensordelimiter{?}
\usepackage[font=scriptsize,labelfont=bf]{caption}
\def\a{\alpha}
\def\b{\beta}
\def\r{\rho}

\def\O{\Omega}

\def\o{\omega}
\def\m{\mu}
\def\g{\gamma}
\def\l{\lambda}
\def\n{\nu}

\def\t{\tau}
\def\s{\sigma}
\def\a{\alpha}
\def\b{\beta}
\def\d{\delta}
\def\g{\gamma}
\def\G{\Gamma}

\def\rd{{\rm d}}

\def\tJ{{\tilde{J}}}

\def\cD{{\mathcal{D}}}
\def\cP{{\mathcal{P}}}
\def\cK{{\mathcal{K}}}
\def\cR{{\mathcal{R}}}
\def\tcR{{\widetilde{\cR}}}
\def\tG{ {\tilde \G}}
\def\tPi{ {\tilde \Pi}}

\def\tkappa{{\tilde{\kappa}}}
\def\tJ{{\tilde{J}}}

\def\TCN{{$\text{TC}\mathcal{N}$}}

\def\be{\begin{equation}} 
\def\ee{\end{equation}}

\begin{document}

\title{General structure of Thomas$-$Whitehead  gravity}

\author{Samuel Brensinger}
 \affiliation{%
 Department of Mathematics\\
The University of Dayton, Dayton, OH 45469, USA}
 \email{sbrensinger1@udayton.edu}

\author{Kenneth Heitritter}
 \affiliation{%
 Department of Physics and Astronomy\\
 The University of Iowa, Iowa City, IA 52242, USA}
 \email{kenneth-heitritter@uiowa.edu}\email{vincent-rodgers@uiowa.edu}

\author{Vincent G. J. Rodgers}
 \affiliation{%
 Department of Physics and Astronomy\\
 The University of Iowa, Iowa City, IA 52242, USA}
 \email{vincent-rodgers@uiowa.edu}

\author{Kory Stiffler}
 \altaffiliation{%
 Department of Physics and Astronomy\\
 The University of Iowa, Iowa City, IA 52242, USA}
 \affiliation{%
 Brown Theoretical Physics Center and Department of Physics\\
 Brown University, Providence, RI 02912-1843, USA}
 \email{kory\_stiffler@brown.edu}

\date{\today}

\begin{abstract}
Thomas-Whitehead (TW) gravity is a projectively invariant model of gravity over a d-dimensional manifold that is intimately related to string theory through reparameterization invariance.  Unparameterized geodesics are the ubiquitous structure that ties together  string theory and higher dimensional gravitation. This is realized through the projective geometry of Tracy Thomas. The projective connection, due to Thomas and later Whitehead, admits a component that in one dimension is in one-to-one correspondence with the coadjoint elements of the Virasoro algebra.   This component is called the diffeomorphism  field $\mathcal{D}_{ab }$ in the literature. It also has  been shown that in four dimensions, the TW\ action collapses to  the    Einstein-Hilbert action with cosmological constant when $\mathcal{D}_{ab}$  is proportional to the Einstein metric.   These previous results have been restricted to either  particular metrics, such as the Polyakov 2D\ metric,  or were restricted to coordinates that were volume preserving.  In this paper, we review TW gravity and derive the    gauge invariant TW action  that is explicitly projectively invariant and general coordinate invariant.  We derive the covariant field equations for the TW  action and show how fermionic fields couple to the gauge invariant theory.       The independent fields are the metric tensor $g_{ab}$, the fundamental projective invariant $\Pi^{a}_{\,\,\,bc}$, and the diffeomorphism field $\mathcal D_{ab}$.  
\end{abstract}

\maketitle

\section{Introduction}
The geometric classification of manifolds via their geodesics as opposed to distances between points (metrical)  is  an old notion.  Indeed in his inaugural professorial lecture at Cambridge University in 1863, Cayley remarked that ``descriptive geometry includes metrical geometry'' and ``descriptive geometry is in fact \emph{all} geometry'' \cite{Crilly1999}. In this reference, descriptive geometry corresponds to  projective geometry. The question of whether the family of geodesics could uniquely determine the metric for general relativity was investigated by Cartan in \cite{Cartan:1923zea,Cartan:1924} and further developed by Thomas \cite{Thomas:1925a,Thomas:1925b} and Whitehead \cite{Whitehead}. The answer is that  connections can only be determined up to equivalences classes. A  resurgence of investigations into the physical ramifications of metrical versus descriptive (projective) geometry can be found in the literature  \cite{Hall:2007wp,Hall:2008,Hall:2009zza,Hall:2011zza,Nurowski:2010wx}.   On the other hand, the Virasoro algebra \cite{Virasoro:1969zu}  is  considered at the heart of  string theory. It is usually viewed  through its relationship with  conformal symmetry, where two copies of the Virasoro algebra define the conformal algebra.  However, the relationship between   string theory and  the Virasoro algebra also has an even more primitive origin through its identity as a one dimensional vector space\cite{Pressley:1988qk} and  projective structure  \cite{Cartan:1923zea,OvsienkoValentin2005Pdgo,Kirillov:1982kav}.  Since the coadjoint orbits admit a natural symplectic structure, their geometric actions provide  an  avenue to the two dimensional field theories that can be associated with quantum gravity \cite{Rai:1989js,Alekseev:1988ce,Alekseev:1988vx}. Furthermore, when married with an affine Lie Algebra (a Kac-Moody algebra), one finds that the coadjoint elements appear as background sources for the two-dimensional gravitation (Virasoro sector) and gauge (Kac-Moody sector) theories.  The background fields in the Kac-Moody sector correspond to the vector potentials which serve as the gauge connections, \( A_{a} \), for Yang-Mills theories.  It was suggested in  \cite{Rodgers:1994ck}  that  the coadjoint elements of the Virasoro sector also could be put on  an equivalent footing with the Kac-Moody sector if the  coadjoint elements of the Virasoro algebra could  also have an  associated ``gauge'' field in higher dimensions. The posited field was dubbed the diffeomorphism field, \( \cD_{ab}\).  This realization was recently established in \cite{Brensinger:2017gtb}, when Kirillov's observation\cite{OvsienkoValentin2005Pdgo,Kirillov:1982kav} that the coadjoint elements of the Virasoro algebra are in one-to-one correspondence with Sturm-Liouville\cite{Liouville:1887,Liouville:1889} operators, was reexamined. The authors were able to   use the one dimensional projective structure to  provide a bridge  between the Virasoro algebra and projective geometry in higher dimensions. Thus the analogous ``gauge'' symmetry due to reparameterization invariance in the Virasoro sector is projective invariance and the  diffeomorphism  field corresponds to projective connections.  With this, the diffeomorphism field that appears in two dimensions through the geometric action as a background field has a different interpretation than that of expectation values of external energy-momentum tensors, as in conformal field theories.  Furthermore the diffeomorphism field can  acquire  dynamics as a fundamental field through the projective curvature squared terms.   Some of the entangled relationship between conformal geometry and projective geometry has been studied in \cite{Nurowski:2010wx,CapA.2014Emip,curry_gover_2018,GoverA.Rod2012DEgE,Gover:2014vxa,BaileyT.N.1994TSBf}. For a good review see \cite{Eastwood2}.   

So far, discussions of dynamical projective connections \cite{Brensinger:2017gtb,Brensinger:2019mnx} have been restricted to particular metrics that are focused on the 2D Polyakov metric \cite{Polyakov:1981rd,Polyakov:1987zb}  or Einstein geometries in four dimensions where compatibility has been enforced.  In this paper we generalize those considerations for any space-time dimensions  and exhibit a Lagrangian that is  explicitly projectively invariant and  general coordinate invariant, i.e. \emph{gauge invariant}.   We will briefly review the salient features of the study of geodesics through the Thomas-Whitehead connection, the Thomas Cone and tensor and fermion representations on the Thomas Cone.  Then, by using the Palatini \cite{palatini} formalism, we explicitly construct the gauge invariant Thomas - Whitehead gravitational action (TW) \cite{Brensinger:2017gtb}, the gauge invariant Dirac action and covariant field equations, its coupling to arbitrary Yang-Mills theories, and the energy-momentum tensor. This work can be extended to include higher-order interactions, using the projective version of Lovelock Gravity\cite{Lovelock:1971yv} to classically maintain an initial value formulation.  We will conclude with remarks on geodesic deviations as it is there that contributions through gravitational radiation may become manifest.       
\section{From Geodesics to  Projective Curvature}
In its most pragmatic form, string theory can be thought of as regulating the Feynman diagrams in gravitational theories by adding a small space-like curve to the point particle.  This activity already endows the string with a projective structure.  The curves are parameterized by vector fields, say \( \zeta^a = \frac{dx^a}{d\sigma} \), which allows one to take the intrinsic or absolute derivative of any vector field along these curves.    In one-dimension the Virasoro algebra is the algebra of centrally extended vector fields on a line or circle and  a projective structure emerges \cite{Cartan:1923zea,Cartan:1924,Kirillov:1982kav}. 
\subsection{Geodetics}
  In any dimension, the intrinsic (or absolute) derivative of a vector field \( v^a \) along a curve $ \mathcal{C}$ parameterized by $\s$ is given by,
\begin{equation} \label{Intrinsic}
\frac{D v^a}{d\sigma} \equiv \frac{d v^a}{d \sigma} + \Gamma^a{}_{bc} v^b \zeta^c,
\end{equation} where \(  \Gamma^a{}_{bc}\) are connection coefficients associated with a connection \(\nabla_a\) and $\zeta^a$ is the tangent vector $\frac{dx^a}{d\s}$ along the curve $\mathcal C$. The connection is assumed to be torsion-free and therefore satisfies the symmetry relation $\Gamma\indices{^a _{bc}}=\Gamma\indices{^a _{cb}}$.
An affine geodesic generalizes the notion of a straight line and \(\zeta^a\) is said to be geodesic  if the change of $\zeta^a$ along the curve $\mathcal{C}  $ parameterized by $\s $ is be proportional to itself, i.e.
\begin{equation}
\frac{D \zeta^a}{d\sigma} = f(\sigma) \zeta^a,
\end{equation}
where $f(\s)$ is the proportionality function.
This yields the affine geodesic equation, 
\begin{equation}
\frac{d^2 x^a}{d \s^2}+ \Gamma^a{}_{bc} \frac{d x^b}{d \s}\frac{d x^c}{d \s} = f(\sigma) \frac{d x^a}{d \s}. \label{affinegeo}
\end{equation}
 One may change the parameterization from $\s$ to $u(\s)$  by writing 
\begin{equation}
\frac{d}{d u} = \frac{d\s}{d u}\frac{d}{d \s}
\end{equation}
      and for a suitable choice $u(\s)$ we can eliminate the right hand side of Eq.~(\ref{affinegeo}) to write the \emph{geodetic} equation
\begin{equation}
\frac{d^2 x^a}{d u^2}+ \Gamma^a{}_{bc} \frac{d x^b}{d u}\frac{d x^c}{d u} =0. \label{affinegeodetic}
\end{equation}
Here the parameter $u$ is said to be an \emph{affine} parameter with respect to the connection $\nabla_a$ as
\begin{displaymath}
 \frac{d x^b}{d u} \nabla_b u = 1.
\end{displaymath}  Although the parameterization may have changed, the curves remain the same. Furthermore different connections, say $\hat \nabla_a$ and $\nabla_a$ can sometimes admit the same geodesics.  If so, then $\hat \nabla$ and $\nabla$ belong to the same projective equivalence class. Thomas showed how one can write a gauge theory over this projective symmetry \cite{Thomas:1925a,Thomas:1925b}. We discuss this gauge theory presently. 

\subsection{Projectively equivalent paths}
    Consider a $\rd$-dimensional manifold $\mathcal{M}$ with coordinates $x^a$ where italic latin indices $a,b,c,m,n,\dots = 0,1,\dots, \rd-1$.  
Let  $\hat{\nabla}_a$ be a connection on $\mathcal{M}$ where  \(\zeta^a \) is geodetic, i.e.
\begin{align}
        \zeta^b \hat{\nabla}_b \zeta^a =\frac{d^2 x^a}{d\tau^2} + \hat\Gamma^a{}_{bc} \frac{d x^b}{d\tau}\frac{d x^c}{d\tau}= 0. & \label{geo1}
\end{align}
Now consider another connection whose coefficients are defined as
\begin{equation}
\Gamma^a{}_{bc} = \hat \Gamma^a{}_{bc} + \delta^a_{\,\,\,b} v_c + \delta^a_{\,\,\,c} v_b, \label{ProjectiveTransformation}
\end{equation}
where $v_b$ is an arbitrary one form. The geodesic equation for this connection is then 
\begin{align}
        \zeta^b {\nabla}_b \zeta^a =\frac{d^2 x^a}{d\tau^2} + \Gamma^a{}_{bc} \frac{d x^b}{d\tau}\frac{d x^c}{d\tau} = f(\tau)\frac{d x^a}{d\tau},  & \label{geo2}
\end{align}
and where \(f(\tau) = 2 v_b \frac{d x^b}{d\tau}.   \)
Since Eq.~(\ref{geo2}) can also be made geodetic by a suitable reparameterization of  $\tau$ to $u(\tau)$ both Eq.~(\ref{geo1}) and Eq.~(\ref{geo2})admit the same geodesic curves.  Eq.~(\ref{ProjectiveTransformation}) is called a \emph{projective transformation} and establishes the projective equivalence relation, $\hat \Gamma^a{}_{bc} \sim  \Gamma^a{}_{bc} $.   

In \cite{Thomas:1925a,Thomas:1925b}, Thomas presents a ``gauge" theory of projectively equivalent connections that is projectively invariant and general coordinate invariant.  This begins by defining the   \emph{fundamental projective invariant} $\Pi^{a}{}_{bc}$ 
\begin{align}\label{e:Pi}
        \Pi^{a}{}_{bc} \equiv \Gamma^a{}_{bc} - \tfrac{1}{(\rd+1)} \delta^a_{\,\,(b}{} \Gamma^m{}_{c)m},
\end{align}
which is traceless by construction
\begin{align}
        \Pi\indices{^a _{ba}}=\Pi\indices{^a _{ab}} = 0
\end{align}
and invariant under a projective transformation,  Eq.~(\ref{ProjectiveTransformation}), for an arbitrary one form $v_a$. Using the fundamental projective invariant $\Pi\indices{^a _{bc}}$ one can write a geodetic equation
\begin{align}
        \frac{d^2 x^a}{d\tau^2} + \Pi^a{}_{bc} \frac{d x^b}{d\tau}\frac{dx^c}{d\tau}=0,
\end{align}
that is projectively invariant.  However this equation is not covariant as $\Pi^{a}{}_{bc} $ transforms as
\begin{eqnarray} 
{\Pi'}^a_{\ bc} = &J^a_{\ f} \left( \Pi^f_{\ de} \bar{J}^d_{\ b} \bar{J}^e_{\ c} + \frac{\partial^2 x^f}{\partial x'^b \ \partial x'^c} \right)\nonumber\\
\label{e:PiTrans} 
 &+ \frac{1}{d+1} \frac{\partial \log |J|}{\partial x^d} \left( \bar{J}^d_{\ b}\delta^a_{\ c} + \bar{J}^d_{\ c}\delta^a_{\ b} \right)  \end{eqnarray}
under a general coordinate transformation from \(x \rightarrow x'(x) \) with \(J^a_{\ b} = \frac{\partial x'^a}{\partial x^b}\), the Jacobian of the transformation. We will denote the  inverse Jacobian as \(\bar{J}^a_{\ b} = \frac{\partial x^a}{\partial x'^b}\).
The last summand  spoils the covariance and can be related to volume, as it involves the determinant of the Jacobian of the transformation \(J=\det{(J^a_{\ b})} \). Thomas then constructs a line bundle over $\mathcal{M}$ which is  a $\rd +1$-dimensional manifold  $\mathcal{N}$ referred to as the Thomas Cone~\footnote{Our extra direction $\lambda$ is related to Thomas' original extra direction in~\cite{Thomas:1925a,Thomas:1925b} through an exponential as in~\cite{Crampin}. Furthermore, Thomas referred to this extra direction as the $0$-th direction where we refer to it as the $\rd$-th direction, reserving the index $0$ for time as is common in the physics literature.} \cite{Eastwood}. The coordinates on the Thomas Cone are  \(( x^0, x^1, \dots , x^{\rd -1} , \lambda  )    \), where $\l$ is denoted the volume coordinate.  Since the  volume coordinate, $\l,$ takes values $0< \l < \infty$,  $\mathcal N$ is called a cone. The coordinates transform as 
\begin{align}\label{e:ConeTrans}
        x'^\a = ( x'^0(x^d), x'^1(x^d), \dots , x'^{\rd -1}(x^d) , \lambda' = \lambda |J|^{-\frac{1}{\rm{d}+1}} ).
\end{align}
From here on, we refer to transformations in Eq.~(\ref{e:ConeTrans}) as $\text{TC}\mathcal{N}$-transformations.
 Here, Greek indices  are over $\mathcal{N}$ coordinates and take values $\alpha, \beta, \mu, \dots = 0,1,2,\dots,\rd$ and italic latin indices are  over coordinates on $\mathcal{M}$ and take values $a,b,m,n,\dots = 0,1,2,\dots \rd-1$. We reserve the index $\lambda$ and the upright letter $\rd$ to refer to the volume coordinate  $x^\rd = x^\lambda =\lambda$. For every coordinate transformation on  $\mathcal{M}$ there is a unique coordinate transformation on  $\mathcal{N}$.

\subsection{Thomas projective connections }
Thomas was able to find a connection on $\mathcal{N}$ that transforms as a connection by extending the fundamental projective invariant to a $\rd+1$-dimensional projective connection $\tPi^{\a}{}_{\m\n}$. It is defined as follows~\cite{Thomas:1925a,Whitehead,Roberts}
\begin{subequations}\label{e:tPi2}
\begin{eqnarray}
&\tPi^{\a}{}_{\lambda \b} &=\tPi^{\a}{}_{\b \lambda} = - \tfrac{1}{\rd+1} \delta^\a_{\,\,\b} \\
&\tPi^a{}_{bc} &=\tPi^a{}_{cb} =\Pi^a{}_{bc} \\
&\tPi^\lambda{}_{ab} &=\tPi^\lambda{}_{ba} = -\tfrac{\rd+1}{\rd-1} \cR_{ab} \; ,
\end{eqnarray}
\end{subequations}
where $\cR_{ab}$ is constructed from the \emph{equi-projective curvature ``tensor''} $\cR^m{}_{abn} $ 
\begin{equation}
~\cR^m{}_{abn} = \Pi^m{}_{a[n,b]} + \Pi^p{}_{a[n}\Pi^m{}_{b]p} \; ,
\end{equation} with an associated \emph{equi-projective Ricci ``tensor"} \begin{align}
        \cR_{ab} = \cR^m{}_{amb}\; . \label{ProjRicci}
\end{align}
With this, $\tPi^\a{}_{\m\n}$ transforms as a connection under a \TCN-transformation as
\begin{align}\label{e:tPiTrans2}
        \tPi'^{\alpha}{}_{\mu\nu} = \frac{\partial x'^\a}{\partial x^\rho}\frac{\partial x^\sigma}{\partial x'^\m}\frac{\partial x^\beta}{\partial x'^\n}\tPi^{\rho}{}_{\sigma\beta} + \frac{\partial^2 x^\beta}{\partial x'^\mu \partial'^\nu}\frac{\partial x'^\a}{\partial x^\b}
\end{align}
so that one may construct a projective curvature tensor
\begin{align}
        \tcR^\a{}_{\m\n\b} =& \tPi^\alpha{}_{\mu[\beta,\nu]} + \tPi^\rho{}_{\mu[\beta}\tPi^\alpha{}_{\nu]\rho} \; , 
\end{align}
whose non-vanishing components are
\begin{align}
\tcR^\lambda{}_{abn} &= \tfrac{\rd+1}{\rd-1} (\cR_{a[b,n]} + \Pi^m{}_{a[b} \cR_{n]m})\\
\tcR^m{}_{abn} &= \cR^m{}_{abn} - \tfrac{1}{\rd-1} \d_{[n}{}^m \cR_{b]a}\;.
\end{align}
The projective Ricci tensor is defined as the trace of the projective curvature tensor and vanishes identically
\begin{align}\label{e:TRicciN2}
        \tcR_{\a\b} \equiv & \tcR^\m{}_{\a\m\b} = 0.
\end{align}
This construction is only a specific example of a projective connection but it laid the ground work for the more general setting we now present.  
\section{Thomas-Whitehead Projective Geometry}
\subsection{The general projective connection}
The original Thomas projective connection,  $\tPi^\a{}_{\m\n}$, can be generalized to a connection ${{\tilde\G}}^{\a}_{\,\,\b \g}$ \cite{Roberts,Crampin,Whitehead}, where explicitly 
\begin{equation} 
{{\tilde\G}}^{\a}_{\,\,\b \g}= \begin{cases}
    {\tilde \G}^{\lambda}_{\,\,\,\lambda a}={\tilde \G}^{\lambda}_{\,\,\, a \lambda} = 0
    \\ {\tilde \G}^{\a}_{\,\,\,\,\lambda \lambda} = 0 \label{e:Gammatilde}\\ {\tilde \G}^{a}_{\,\,\,\,\lambda b}={\tilde \G}^{a}_{\,\,\,\,b \lambda} = \a_\lambda\,\d^a_{\,\,b}\\
{\tilde \G}^{a}_{\,\,\,\,b c} ={ \Pi}^{a}_{\,\,\,\,b c}\\
{\tilde \G}^{\lambda}_{\,\,\,\, a b} =  \Upsilon^\lambda \cD_{ a b}
 \end{cases} 
\end{equation} 
and where 
\begin{align}
{\Pi}^{a}_{\,\,\,\,b c} &= { \G}^{a}_{\,\,\,\,b c} + \delta^a{}_{(c}~ \a_{b)}\\
\a_a &= -\tfrac{1}{\rm{d} +1} \Gamma^m{}_{am}\\
\Upsilon^\a &= (0,0,\dots,0,\lambda)\\
\a_\a &= \left( \a_a, \lambda^{-1} \right) \;. \label{UpsilonOmega}
\end{align}
Here the connection ${ \G}^{a}_{\,\,\,\,b c} $ is \emph{any} representative member of the equivalence class $[{ \G}^{a}_{\,\,\,\,b c}]$ of projectively equivalent connections, related via Eq.~\ref{ProjectiveTransformation}, and $\a_a$ is that chosen member's trace component. However, keep in mind that   \({\Pi}^{a}_{\,\,\,\,b c}  \) exists in its own right in that it is traceless and transforms like a traceless part of an affine connection. Notice also  that only the $\lambda$ component for $\a_\m$ appears in the projective connection $\tilde \G^{\m}_{\,\,\,\a \b}$. On $\mathcal{M ,}$  the transformation laws are
\begin{align}\label{GTrans}
        {\G}'^{a}{}_{mn} &= \frac{\partial x'^a}{\partial x^b}\frac{\partial x^p}{\partial x'^m}\frac{\partial x^q}{\partial x'^n}{\G}^{b}{}_{pq} + \frac{\partial^2 x^b}{\partial x'^m \partial'^n}\frac{\partial x'^a}{\partial x^b}~~~\\
        \a'_a &= \frac{\partial x^m}{\partial x'^a} \a_m +  \frac{\partial \log  |J|^{\frac{1}{\rm{d}+1}} }{\partial x'^a}.\label{omegaTrans}
\end{align} 
In the above, $\cD_{ab}$ generalizes the work of Thomas and transforms in such a way that ${{\tilde\G}}^{\a}_{\,\,\b \g}$ transforms as an affine connection on $\mathcal{N}$. This is the origin of the diffeomorphism field $\mathcal{D}_{ab}$.  In this construction,  $\Upsilon$ is the fundamental vector on the Thomas cone and satisfies the compatibility relation
\be {\tilde \nabla}_\a \Upsilon^\b = \d_\a{}^\b \; ,\ee so $\Upsilon^\b $ satisfies the fundamental geodesic equation with unit proportionality
\be \Upsilon^\b {\tilde \nabla}_\b \Upsilon^\a =  \Upsilon^\a\; . \ee
For functions on $\mathcal{N}$
\begin{equation}
\Upsilon^\b {\tilde \nabla}_\b  f= \l \partial_\l f \;, 
\end{equation} 
showing that $\Upsilon$ generates scaling in the $\lambda$ direction. One-forms $\b_\a$ on $\mathcal{N}$ are uniquely defined by $\b_a$ on $\mathcal M$ when   $\b_\a \Upsilon^\a=1  $ and  the Lie derivative with respect to $\Upsilon $  vanishes i.e, $\mathcal{L}_\Upsilon \b_\rho =0,$ so that it is scale invariant.
 Under a \TCN-transformation, Eq.~(\ref{e:ConeTrans}), $\Upsilon^\a$ and the covariant derivative transform as
\begin{align}
        \Upsilon'^\a &= \frac{\partial x'^\a}{\partial x^\b} \Upsilon^\b \\
        \nabla'_\a &= \frac{\partial x^\b}{\partial x'^\a} \nabla_\b\;.
\end{align}
Demanding that $\tG^\a{}_{\m\n}$  transforms as an affine connection
 \begin{align}\label{e:tGTrans}
        \tG'^{\alpha}{}_{\mu\nu} = \frac{\partial x'^\a}{\partial x^\rho}\frac{\partial x^\sigma}{\partial x'^\m}\frac{\partial x^\beta}{\partial x'^\n}\tG^{\rho}{}_{\sigma\beta} + \frac{\partial^2 x^\beta}{\partial x'^\mu \partial'^\nu}\frac{\partial x'^\a}{\partial x^\b},
\end{align}
and using the transformation laws  of $\Upsilon^\a$ and $\tG^\lambda{}_{ab},$  one finds that $\cD_{ab}$ transforms under a coordinate transformation on $\mathcal{M}$ as
\begin{align}
        \cD'_{ab} = \frac{\partial x^m}{\partial x'^a}\frac{\partial x^n}{\partial x'^b} (\cD_{mn} -\partial_m j_n-j_m j_n+j_c\Pi^c{}_{mn}),\label{DiffTransformation} 
\end{align}
 where we define $j_a = \partial_a \log{|J|^{-\frac{1}{\rm{d}+1}}}$. One can show that the coordinate transformation law of \(\mathcal{D}_{ab}\) as stated by Eq.~(\ref{DiffTransformation}) is an action of the general linear group on the components of \(\mathcal{D}\). This property holds despite the presence of the coordinate-dependent object \(\Pi^a_{\ bc}\) in the transformation law\cite{SamBrensinger}. This transformation law will become important later in the correspondence with coadjoint elements of the Virasoro algebra in one-dimension. 

A general tensor on $\mathcal{M}$ with $m$-contravariant and $n$-covariant indices we express as
\begin{align}
T^{a(m)}{}_{b(n)} = T^{a_1 a_2 \dots a_m}{}_{b_1 b_2 \dots b_n}\;.
\end{align}In what follows we refer to  $(m,n)$-tensor on $\mathcal{M}$ as objects that transform as 
\begin{align}\label{e:MTensTrans}
        T'^{a(m)}{}_{b(n)} = \tfrac{\partial x'^{a_1}}{\partial x^{p_1}} \dots \tfrac{\partial x'^{a_m}}{\partial x^{p_m}}\tfrac{\partial x^{q_1}}{\partial x'^{b_1}}\dots \tfrac{\partial x^{q_n}}{\partial x'^{b_n}}T^{p(m)}{}_{q(n)} 
\end{align}
under coordinate transformations. Similarly,  we refer to  objects as $(m,n)$-TC tensors on $\mathcal{N }$ that transform as \begin{align}\label{e:PTensTrans}
        T'^{\a(m)}{}_{\b(n)} = \tfrac{\partial x'^{\a_1}}{\partial x^{\m_1}} \dots \tfrac{\partial x'^{\a_m}}{\partial x^{\m_m}}\tfrac{\partial x^{\n_1}}{\partial x'^{\b_1}}\dots \tfrac{\partial x^{\n_n}}{\partial x'^{\b_n}}T^{\m(m)}{}_{\n(n)}
\end{align} under a \TCN-transformation. This will allow us to build actions that are invariant with respect to \TCN-transformations. \subsection{Geodetics revisited}
Before discussing projective curvature relations, we now revisit geodesics and geodetics to illuminate the projective connection.  Consider a geodetic on $\mathcal{N}$ associated with the vector field \( \zeta^\a = \frac{d x^\a}{du}\).  The parameter $u$ is an affine parameter for $\tilde \nabla$ such that
\begin{equation}
\zeta^\a \tilde \nabla_\a \zeta^\b =0 \;.
\end{equation}
Separating the $\mathcal{M}$ coordinates from $\l$, we have the expressions
\begin{eqnarray}
&&\frac{d^2 x^a}{du^2}+ { \Pi}^{a}_{\,\,\,\,b c}\frac{dx^b}{du}\frac{dx^c}{du} = -2 \frac{1}{\l} \left(\frac{d\l}{du}\right)\frac{dx^a}{du}, \label{Aeq} \\
&& \frac{d^2 \l}{du^2} + \l \cD_{bc} \frac{dx^b}{du}\frac{dx^c}{du}=0.\label{Beq}
\end{eqnarray} 
Together, these equations are covariant and projectively invariant.  Let us consider a reparameterization that can render Eq.~(\ref{Aeq}) geodetic.  In other words, does there exist a parameter $\t$ that is affine with respect to the projective invariant ${ \Pi}^{a}_{\,\,\,\,b c} $? Let $u \rightarrow \tau(u)$ so that 
\begin{equation}\label{u goes to tau}
\frac{d^2 \t}{du^2} = -2\left( \frac{1}{\l} \frac{d\l}{du} \right)\frac{d\t}{du}\;.
\end{equation} 
This will eliminate the RHS of Eq.~(\ref{Aeq}) and we can use this to eliminate  $\l$ in Eq.~(\ref{Beq}) with 
\begin{equation}
\frac{d^2 \l}{du^2} = \frac{\l}{4}\cdot\frac{3 (\frac{d^2 \t}{du^2})^2- 2 (\frac{d^3\t}{du^3}) \frac{d\t}{du}}{ (\frac{d\t}{du})^2}\;.
\end{equation}
With this, one finds that the reparameterization is viable if
\begin{equation}
 \cD_{bc} \frac{dx^b}{du}\frac{dx^c}{du} = \frac{1}{2}\cdot \frac{\frac{d\t}{du} (\frac{d^3 \t}{du^3})- \frac{3}{2} (\frac{d^2 \t}{du^2})^2}{ (\frac{d\t}{du})^2}\equiv \frac{1}{2} S(\t:u)\;, \label{Schwartzian Derivative}
\end{equation}  
where $S(\t:u)$ is the Schwarzian derivative of $\t$ with respect to $u$.  For example, if the kinetic term $\cD_{bc} \frac{dx^b}{du}\frac{dx^c}{du}$ vanishes, then requisite reparameterizations that  render $\t$ affine  are the  M\"obius transformations $\t = \frac{a u + b}{c u +d} $, where $a,b,c,$ and $d$ are real numbers.  Another familiar example is when   $\cD_{bc} \frac{dx^b}{du}\frac{dx^c}{du} = \frac{m^2-1}{2 u^2} $  and the requisite transformation are the exponential M\"obius transformations  $\t = (\frac{a u^m + b}{c u^m +d})$.        This corresponds to the coadjoint orbits of the Virasoro algebra denoted by $\text{Diff}(S^1)/\text{SL}(2,m)$, where the isotropy group is generated by $L_m, L_0, L_{-m}$. A M\"obius transformation is a one-dimensional projective transformation, so we see that the preferred class of parameters for \(\Pi^a_{\ bc}\) is preserved by projective transformations rather than affine transformations. This motivates the description of \(\Pi^a_{\ bc}\) as a \textit{projective connection}. The inclusion of \(\Pi^a_{\ bc}\) in the TW connection, which incorporates the field \(\mathcal{D}_{bc}\), allows us to apply techniques that are typically available for affine connections.

\subsection{Projective geometry}
One constructs the projective curvature tensor in the usual way 
 \label{e:Kaction}\begin{align} 
[{\tilde \nabla}_\a,{\tilde \nabla}_\b] V^\g &= \,{\cK}^{\g}_{\,\,\,\rho\a \b  } V^\rho \\
[{\tilde \nabla}_\a,{\tilde \nabla}_\b] V_\g &=- \,{\cK}^{\rho}_{\,\,\,\g\a \b  } V_\rho \; .
\end{align} 
from connections that transform as in  Eq.~(\ref{e:tGTrans}). In terms of the connections, the curvature can be written explicitly as
\begin{align}
        \cK^\a{}_{\m\n\b} = \tG^\a{}_{\m[\b,\n]} + \tG^\rho{}_{\m[\b}\tG^{\a}{}_{\n]\rho} \;.
\end{align}
This transforms as a (1,3) TC tensor on $\mathcal{N}$. Using Eq.~(\ref{e:Gammatilde}) to  expand $\tilde{\Gamma}^{\a}{}_{\m\n}$ we find the only non-vanishing components of the \emph{projective curvature tensor} to be
\begin{align} \label{eq:Projective Curvature Tensor Components} \begin{split} \cK^{a}_{\ bcd} &= \mathcal{R}^{a}_{\ bcd} + \delta^a_{\ [c}\mathcal{D}_{d]b} \\ \cK^{\lambda}_{\ cab} &= \lambda \partial_{[a}\mathcal{D}_{b]c} + \lambda \Pi^{d}_{\ c[b}\mathcal{D}_{a]d}. \end{split} \end{align}
We will also find it useful later on to have a $\lambda$-independent version of $\mathcal{K}\indices{^\lambda _{cab}}$. We define this symbol as
\begin{align}
\label{K3_nolambda}
\breve{\mathcal{K}}\indices{_{cab}}\equiv \frac{1}{\lambda}\mathcal{K}\indices{^\lambda _{cab}}=\partial_{[a}\mathcal{D}_{b]c} + \Pi^{d}_{\ c[b}\mathcal{D}_{a]d}\;.
\end{align}

By contracting the first and third indices of the projective curvature tensor, we can write the  projective Ricci tensor whose only non-vanishing components are
\begin{equation} \label{eq:Projective Ricci} \mathcal{\cK}_{bd} = \mathcal{R}_{ bd}+(\rd-1) \cD_{bd}.\end{equation}
$ \mathcal{R}_{bd} $ is the equi-projective Ricci tensor from Eq.~(\ref{ProjRicci}).  The expressions in Eq.~(\ref{eq:Projective Curvature Tensor Components}) are precisely of the form seen in conformal geometry 
\begin{align} \label{eq:Conformal Curvature Tensor Components} \begin{split} R^{a}_{\ bcd} &= W^{a}_{\ bcd} + \delta^a_{\ [c}{P}_{d]b} \\ C_{ cab} &= \partial_{[a}{P}_{b]c} +  \G^{d}_{\ c[b}{P}_{a]d}\; , \end{split} \end{align}
where $W^{a}_{\ bcd}$ is the Weyl tensor, ${P}_{d b}$ is the Schouten tensor, and $C_{\ cab}$ is the Cotton-York tensor. In the above, $W^{a}_{\ bcd} $ is analogous to $\cK^{a}_{\ bcd}$ in Eq.~(\ref{eq:Projective Curvature Tensor Components}). If we consider the contraction of the projective curvature tensor with a volume one-form $g_\mu,$ that transforms as Eq.~(\ref{omegaTrans}) and is also invariant under projective transformations, we can form  the \emph{projective Cotton-York} tensor, ${\mathcal{K}(g)}_{\n \a\b}\equiv g_\mu \mathcal{K}^\mu{}_{\n \a\b}$. Then we can write
\begin{align}
 \mathcal{K}(g)_{nab} &=g_\mu \mathcal{K}^\mu{}_{nab} \nonumber\\
 =&\mathcal{P}_{[b|n|;a]}- \Delta_{n}\mathcal{P}_{[a b]} + \Delta_{[a}\mathcal{P}_{b]n} + R^{m}{}_{nab}\Delta _m,
\end{align}
where $\Delta_a\equiv g_a -\a_a $ is a one form on $\mathcal{M}$. $\mathcal{K}(g)_{\n \a\b} $ is now explicitly seen as a (0,3)-TC tensor on $\mathcal{N}$ and $\mathcal{K}_{nab} $ is a (0,3)-tensor on $\mathcal{M}$. When we introduce a metric tensor $g_{am}$ on $\mathcal M$ in the next section, we will find that $g_\m=(g_a, \frac{1}{\l})$, where \(g_a \equiv -\frac{1}{\rm{d}+1}\,\partial_a \log \sqrt{|g|}\), is a suitable volume one-form, Eq.~(\ref{eq:Big Metric}). This also introduces the \emph{projective Schouten  tensor} \cite{Gover:2014vxa} \(\mathcal{P}_{ab}\), which is a (0,2)-tensor on $\mathcal{M}$.  The form of $\tG^\a{}_{\m\n}$ in Eq.~(\ref{e:Gammatilde}) allows for  $\cD_{ab}$ to become dynamical as $\cK^\a{}_{\m\a\b} \ne0$, relaxing the Ricci flat condition in ~\cite{Thomas:1925a,Thomas:1925b,Crampin}.  This allows us to extend the Einstein-Hilbert action to projective geometry  as in~\cite{Brensinger:2017gtb,Brensinger:2019mnx}.

If we choose a  member of  the equivalence class $[\G^c{}_{ab}]$,  then we may express $\Pi^c{}_{ab}$ in terms of a specific connection and its associated trace $\a_\m$.  With this, one may write $\cP_{a b}$ in terms of $\cD_{a b} $  as 
\begin{align}
                \mathcal{P}_{bc} = \mathcal{D}_{bc} - \partial_b \a_c + \Gamma^e_{\ bc}\a_e + \a_b \a_c\;.  \label{P_in_terms_of_D}
\end{align} 
   The above is a generalization of~\cite{Brensinger:2017gtb,Brensinger:2019mnx}, where constant volume  coordinates were used and \(\Gamma^e_{\ bc}\) was regarded as Levi-Civita so  $\a_a =0$. Then, in that case,  $\cD_{ab} = \cP_{ab}$ and is a tensor in the volume preserving coordinates.  
As stated above,  $\mathcal{P}_{ab}$ transforms as a tensor on $\mathcal{M}$
\begin{align}\label{e:DTrans}
        \cP'_{ab} = \frac{\partial x^{m}}{\partial x'^{a}}\frac{\partial x^n}{\partial x'^{n}}\cP_{mn},
\end{align} which we may call the projective Schouten tensor in analogy with conformal geometry.

\section{Covariant Metric Tensor on \texorpdfstring{$\mathcal{N}$}{N}}

In projective geometry, a vector field  $\chi$ on $\mathcal{M}$ may be lifted  to a vector field $\tilde \chi$ on $\mathcal{N}$ by writing
\begin{equation}\tilde \chi^\a \partial_\a =-(\l \,\chi^a  \kappa_a {)\partial_\l + \chi^a \partial_a}\;, \end{equation} where $\kappa_a $ is some object that transforms as $j_a$ in Eq.~(\ref{DiffTransformation}), i.e.  
\begin{equation}
\kappa'_a = \frac{\partial x^m}{\partial x'^a} \kappa_m - \frac{1}{\rd+1} \frac{\partial \log J}{\partial x'^a}
\end{equation} under a general coordinate transformation on $\mathcal{M}$. We write the components of $\tilde \chi$ as
 \begin{equation}\tilde \chi^\a = (\chi^a, -\l x^b \kappa_b).\label{VectorOnN}  \end{equation}
  Similarly, if a one form  $v$ on $\mathcal M$ can be related to a projective one form $\tilde v$ via  
 \begin{equation}  \tilde v_\b = (v_b+ \kappa_b, \frac{1}{\l})\;. \label{OneFornOnN}\end{equation}
  It is clear that $\tilde \chi^\a \tilde v_\a = \chi^a v_a $.  A generic vector on $\mathcal{N}$, which has components $\eta_\perp$ that are unrelated to vectors on $\mathcal{M}$, may be written as
 \be\tilde \eta^\b = \left(\eta_\parallel^b,\l( \eta_\perp - \kappa_a \eta_\parallel^a) \right)\;. \label{General Projected Vector}\ee The fundamental vector field $\Upsilon$ in Eq.~(\ref{UpsilonOmega}) has no component parallel to $\mathcal M$, for example.  

We are interested in building an invariant action using the projective curvature.  This will require  a soldering metric which transforms as a tensor on $\mathcal N$ and which is projectively invariant. Taking a metric $g_{ab}$ on $\mathcal M$, one may view this soldering metric as the local tensor product of two one-forms and write
 \begin{equation}\label{eq:Big Metric}
  G_{\mu \nu} = \begin{bmatrix}
  g_{ab}-\lambda_0^{\ 2}g_ag_b & -\frac{\lambda_0^{\ 2}}{\lambda} g_a \\
  -\frac{\lambda_0^{\ 2}}{\lambda} g_b & -\frac{\lambda_0^{\ 2}}{\lambda^2}
  \end{bmatrix}\;.  \end{equation}
Here we have replaced $\kappa_a$ with \(g_a \equiv -\frac{1}{\rm{d}+1}\,\partial_a \log \sqrt{|g|}\) as it is naturally built from the metric degrees of freedom and does not introduce a connection.   The constant \(\lambda_0\) has units of length (like \(\lambda\)), and ensures that \(G_{\mu \nu}\) remains dimensionless when $g_{ab}$ is dimensionless. Since \(G_{\mu \nu}\) depends only on the spacetime metric \(g_{ab}\), it is indeed projectively invariant. One can check that \(G_{\mu \nu}\) satisfies the transformation law
\begin{equation} \label{eq:Transformation of Big G} G'(y)_{\mu \nu} = \frac{\partial x^{\alpha}}{\partial y^{\mu}}\frac{\partial x^{\beta}}{\partial y^{\nu}}G(x)_{\alpha \beta} \end{equation}
when \((x^a,x^{\lambda}) \to (y^a,y^{\lambda}) = (y^a,x^{\lambda}|J|^{-\frac{1}{\rm{d}+1}})\).
Furthermore, under this coordinate  change the volume form on $\mathcal{N } $ remains invariant, i.e.  
\begin{equation} \label{eq:Change of Measure on VM} \sqrt{|G(x^a,x^{\lambda})|} \ dx^{\lambda} d^{\rm{d}}x = \sqrt{|G(y^a,y^{\lambda})|} \ dy^{\lambda} d^{\rm{d}}y \;. \end{equation}
Here \(G(x^a,x^{\lambda})\) and \(G(y^a,y^{\lambda})\) are the metric determinants in the different coordinates. This follows since from Eq.~(\ref{eq:Big Metric}), we see that 
\begin{equation} \label{eq:Relation Between Metric Determinants} |G| = |g|\cdot \frac{\l^2_0}{\l^2} \;, \end{equation}
where \(g\) is the determinant of \(g_{ab}\) on \(M\). Since   \(y^{\lambda} = x^{\lambda}|J|^{-\frac{1}{\rm{d}+1}}\) and \(\frac{1}{\lambda} \to \frac{1}{\lambda} |J|^{\frac{1}{\rm{d}+1}},\) these terms exactly conspire in Eq.~(\ref{eq:Change of Measure on VM}) to maintain the invariant volume on $\mathcal{N}$.  Again, this  motivates why $\l$ is called the volume coordinate. Lastly, the inverse of \(G_{\mu \nu}\) is given by
\begin{equation}\label{eq:Big Metric Inverse}
  G^{\mu \nu} = \begin{bmatrix}
  g^{ab} & -\lambda g^{am}g_m \\
  -\lambda g^{bm}g_m & \frac{\lambda^2}{\lambda_0^{\ 2}}\left(-1 + g^{mn}\lambda_0^{\ 2}g_mg_n\right)
  \end{bmatrix} \;, \end{equation}
 where \(g^{ab}\) is the inverse of the spacetime metric \(g_{ab}\).
 This metric generalizes the work in \cite{Brensinger:2017gtb,Brensinger:2019mnx}, allowing TW gravity to be used in any coordinates.  
We can succinctly write the metric and its inverse as
\begin{align}\label{e:BigGSuccinct}
        G_{\a\b} &= \delta^a_{\,\,\alpha} \delta^b_{\,\,\beta} \,g_{ab} - \lambda_0^2 g_\alpha g_\b\\
        G^{\a\b} &= g^{ab} (\delta^\alpha_{\,\,a} - g_a \Upsilon^\a)(\delta^\b_{\,\,b} - g_b \Upsilon^\b) - \lambda_0^{-2} \Upsilon^\a\Upsilon^\b,
\end{align} 
where we have defined $g_\a \equiv (g_a, \frac{1}{\l})$.
 In TW gravity,  the metric $g_{ab}$, the projective invariant $\Pi^a_{\,\,bc}$, and the diffeomorphism field $\cD_{ab}$ will be treated as independent degrees of freedom in the spirit of the Palatini formalism \cite{palatini}.
\section{ \texorpdfstring{$\tilde \g^\m$ on $\mathcal{N}$}{The Metric and Gamma Matrices on N} }

 Now we seek the $\tilde\g^\a$ matrices associated with the projective metric \(G_{\mu \nu}\) given by Equation \ref{eq:Big Metric}. The gamma matrices, \(\gamma^{m}\), on a \(\rd\)-dimensional spacetime are defined by 
 \begin{equation} \label{eq:Gamma Matrices Definition} \{ \gamma^{m},\gamma^{n} \} = 2 g^{m n} I_{N} \;, \end{equation}
 where \(\{\cdot,\cdot\}\) is the anti-commutator, \(g_{\mu \nu}\) is the spacetime metric, \(N = 2^{\lfloor\rm{d}/2 \rfloor}\), and \(I_{N}\) is the \(N \times N\) identity matrix.

Let \(\tilde{\gamma}^{\mu}\) be the gamma matrices for the metric \(G_{\mu \nu}\) on \(\mathcal{N}\). These matrices satisfy
 \begin{equation} \label{eq:Big Metric Gamma Matrices Definition} \{ \tilde{\gamma}^{\mu},\tilde{\gamma}^{\nu} \} = 2 G^{\mu \nu} I_{N} \end{equation}
 as in Eq.~(\ref{eq:Gamma Matrices Definition}). We will stay in even space-time dimensions. In this case, the gamma matrices \(\tilde{\gamma}^{\mu}\) for \(G_{\mu \nu}\) will have the same dimension as the gamma matrices \(\gamma^{m}\) for \(g_{m n}\).

 Using the inverse of \(G_{\mu \nu}\), Eq.~(\ref{eq:Big Metric Inverse}), we immediately must have \(\tilde{\gamma}^{\mu}=\gamma^{\mu}\) if \(\mu\) is a spacetime coordinate index, say $m$, and where \(\gamma^{m}\) are the gamma matrices for the spacetime metric \(g_{m n}\). The remaining gamma matrix is \(\tilde{\gamma}^{\lambda}\). This matrix must satisfy
 \begin{align} \label{eq:Extra Gamma Matrix Conditions1}
\{ \tilde{\gamma}^{\lambda},\tilde{\gamma}^{m} \} &= -2\lambda g^{m n}g_{n}I_{N}\;, \ \ \  \ \ \ m=0,\dots,\rm{d}-1\;, \\
\label{eq:Extra Gamma Matrix Conditions2}
 2\left(\tilde{\gamma}^{\lambda}\right)^2 &= \{ \tilde{\gamma}^{\lambda},\tilde{\gamma}^{\lambda} \} = 2 \frac{\lambda^2}{\lambda_0^{\ 2}} (-1+g^{m n}\lambda_0^{\ 2}g_{m}g_{n})I_{N}\;. 
\end{align}
  Recall the chiral matrix \(\gamma^5\) in  four-dimensional spacetime.  We will refer to it as  \(\gamma^{\rm{d}+1}\) in the general even dimensional case. It satisfies
 \begin{align} \label{eq:Gamma 5 Conditions1} 
\{ \gamma^{\rm{d}+1},\gamma^{m} \} &= 0 \\ 
\label{eq:Gamma 5 Conditions2} \left( \gamma^{\rm{d}+1} \right)^2 &= I_N\;. 
\end{align}
 Comparing Eqs.~(\ref{eq:Extra Gamma Matrix Conditions1}) and (\ref{eq:Extra Gamma Matrix Conditions2}) to Eqs.~(\ref{eq:Gamma 5 Conditions1}) and (\ref{eq:Gamma 5 Conditions2}), we see that we should have
 \begin{equation} \label{eq:Extra Gamma Matrix} \tilde{\gamma}^{\lambda} = - \frac{\lambda}{\lambda_0} \left( i \gamma^{\rm{d}+1} + \lambda_0g_{m}\gamma^{m} \right) \end{equation}
 as the final gamma matrix for \(G_{\mu \nu}\). Explicitly, the chiral gamma matrix \(\gamma^{\rm{d}+1}\) has the following construction in terms of the other gamma matrices in $\rd$-dimensions
 \begin{equation} \label{eq:Gamma 5 Definition} \gamma^{\rm{d}+1} = \frac{i^{\frac{\rm{d}-2}{2}}}{\rd!} \epsilon_{a_1 \dots a_{\rm{d}}}\gamma^{a_1}\dots \gamma^{a_{\rm{d}}}\;, \end{equation}
 where \(a_i = 0,\dots,\rm{d}-1\) and \(\epsilon\) is the totally antisymmetric Levi-Civita tensor on $\mathcal M$. Specifically, for \(\rm{d}=4\), the gamma matrices for \(G_{\mu \nu}\) are
 \begin{align} \label{eq:VM Gamma Matrices in 4D} \begin{split} \tilde{\gamma}^{m} &= \gamma^{m} \ \ \ \text{when} \ \ \ m=0,1,2,3 \\ \tilde{\gamma}^{\lambda} &= - \frac{\lambda}{\lambda_0} \left( i \gamma^5 + \lambda_0g_{m}\gamma^{m} \right) \;. \end{split} \end{align}
 The fifth gamma matrix \(\gamma^5\) is crucial in discussions about chirality, which we will see when we apply the TW connection to spinor fields. Eq.~(\ref{eq:VM Gamma Matrices in 4D}) shows that the volume bundle metric \(G_{\mu \nu}\) explicitly builds in \(\gamma^5\). Thus, we will expect our dynamical theory for \(\mathcal{D}_{m n}\) to be chiral in nature when interacting with fermions.

 Eqs.~(\ref{eq:Extra Gamma Matrix}, \ref{eq:Gamma 5 Definition},  \ref{eq:VM Gamma Matrices in 4D}) also serve to further establish the relationship between the projective gauge field \(\mathcal{D}_{m n}\) and the notion of volume on \(\mathcal{M}\). Any Lagrangian for \(\mathcal{D}_{m n}\) will involve the metric \(G_{\mu \nu}\) on \(\mathcal{N}\), which in turn can be constructed from gamma matrices. Eq.~(\ref{eq:Extra Gamma Matrix}) says that one of these gamma matrices includes a rescaling of \(\gamma^{\rm{d}+1}\) by \(\lambda\), where \(\gamma^{\rm{d}+1}\) is itself related to volume due to the presence of the epsilon tensor \(\epsilon_{a_1 \dots a_{\rm d}}\). The epsilon tensor is alternating in its indices and transforms as a tensor density that is used to construct volume forms on \(\mathcal{M}\). Therefore, we can again, view \(\lambda\) as a parameter which determines a rescaling of the volume element on \(\mathcal{M}\).
\section{The Virasoro Algebra and Projective Geometry}
Here we will review  three ways in which there is a correspondence between the projective connection's reduction to one dimension and the coadjoint elements of the Virasoro algebra. The Virasoro algebra \cite{OvsienkoValentin2005Pdgo,Segal:1981ap,Witten:1987ty} may be regarded as the centrally extended algebra of vector fields in one dimension.  Let $(\xi,a)$ and \(({\eta},b)\) denote  centrally extended vector fields in one dimension where $a $  and $b$ are  elements in the center.  Then the Lie algebra of these centrally extended vector fields is given through the commutator
\be 
[({\xi}, a), ({\eta},b) ] = ({\xi \circ \eta}, ((\xi,\eta))_{0})\;, 
\ee
where $\xi \circ \eta$ is defined via 
\begin{equation}
\xi \circ \eta \equiv \xi^a \partial_a \eta^b - \eta^a \partial_a \xi^b.
\end{equation} Here we explicitly expose the valence of the one dimensional vectors.  The symbol $((\xi,\eta))_{0} $ is called the Gelfand-Fuchs two-cocycle \cite{Gelfand} and is defined explicitly as 
\begin{align}
\label{Gelfand-Fuchs1} 
((\xi,\eta))_{0} &\equiv
 \frac{c}{2\pi} \int (\xi \eta''' )\,d\theta\\ &=\frac{c}{2\pi} \int \xi^a\nabla_a (g^{bc}\nabla_b \nabla_c \eta^m)g_{mn}d\theta^n \;,  
\label{Gelfand-Fuchs2} 
\end{align}where $g_{ab}$ is a one-dimensional metric.  Eqs.~(\ref{Gelfand-Fuchs1}) and (\ref{Gelfand-Fuchs2}) demonstrate an invariant pairing between $\xi$ and $\eta'''$.  The Gelfand-Fuchs two-cocycle is an example of an invariant pairing between a vector and a quadratic differential $B$ 
\be  < (\xi,a) | (B,c) > \equiv \int (\xi B) d \theta+  ac= \int (\xi^i B_{ij}) d \theta^j +ac\;. \label{pair} \ee In the Gelfand-Fuchs two-cocycle,  the pairing is between a  vector $\xi$ and  a one-cocycle of $\eta,$ where this one-cocycle is a  projective transformation\cite{Kirillov:1982kav,OvsienkoValentin2005Pdgo} that has mapped the    vector field $\eta$  into a quadratic differential. Explicitly,\begin{equation}
\eta \partial_\theta  \rightarrow \eta''' d\theta^2= \nabla_a (g^{bc}\nabla_b \nabla_c \eta^m)g_{mn}d\theta^a d\theta^n. \label{mapping}
\end{equation} The invariant pairing in Eq.~(\ref{pair}) follows if the action of another centrally extended algebra element, say $(\eta, d)$,  leaves the pairing invariant, i.e. 
\begin{equation}
(\eta, d)\ast< (\xi,a) | (B, c) > =0.\end{equation}
This defines the coadjoint representation of the Virasoro algebra \cite{Kirillov:1982kav,Witten:1987ty}.    
\be
ad^*_{(\eta,d)}( B, c)=(\eta B' + 2 \eta' B -  c\, \eta''',0)\;.
 \label{coadjointaction}
\ee
Then, a more general  invariant two-cocycle relative to  the centrally extended coadjoint element $\mathcal{B}=( B, c) $ can be written as
\begin{equation}
(\xi,\eta)_{(B, c)} = \frac{ c}{2\pi} \int (\xi \eta''' - \xi'''
  \eta)\,dx
+ \frac{1}{2\pi} \int (\xi \eta' - \xi' \eta )B\,dx\;.
\label{2cocycle2}
\end{equation}
 One sees that the Gelfand-Fuchs case lives in the pure gauge sector, i.e. $\mathcal{B}=(0,c)$, of the space of coadjoint elements. It was also observed \cite{Kirillov:1982kav} that this action is the same as the action of the space of Sturm-Liouville operators on vector fields.   Thus there is a  correspondence
\be 
(B, c) \Leftrightarrow -2 c \frac{d^2}{dx^2} + B(x)\;,
\label{correspondence}\ee
where on the left side $(B, c)$ is identified with a centrally extended coadjoint element of the Virasoro algebra and on the right side  is a Sturm-Liouville operator with weight $c$ and $B(x)$ as the  Sturm-Liouville potential. 
\subsection{Correspondence through the transformation laws}
Here, we show how the relation between a coadjoint element of the Virasoro algebra and the Sturm-Liouville operator is reconciled by Thomas-Whitehead projective connections.  We will evaluate the connection in one-dimension where one can construct a Laplacian even though curvature is unavailable.  

Consider the transformation of the diffeomorphism field $\cD_{ab}$ in one dimension.  One can show that in one dimension, Eq.~(\ref{DiffTransformation}), i.e.
 \begin{align}
        \cD'_{ab} = & \frac{\partial x^m}{\partial x'^a}\frac{\partial x^n}{\partial x'_b} \cD_{mn} - \frac{1}{(\rd+1)^2}\frac{\partial \log J}{\partial x'^a}\frac{\partial \log J}{\partial x'^b}\nonumber\\
 &- \frac{1}{\rd+1} \frac{\partial^2 \log J}{\partial x'^a \partial x'^b} + \frac{1}{\rd+1} \frac{\partial \log J}{\partial x'^c} \Pi'^c{}_{ab}, 
\end{align}
reduces to\cite{SamBrensinger} 
\begin{equation} \label{eq:1D Infinitesimal Transformation of P 2} \delta \mathcal{D} = 2\xi'\mathcal{D} + \mathcal{D}'\xi - \frac{1}{2}\xi'''  \end{equation}
under an infinitesimal coordinate transformation.   We may let \(\mathcal{D} = qD\) where \(q\) is an arbitrary constant. Then \begin{align} \label{eq:Diffeomorphism Field as Projective Gauge Field} \begin{split} \delta (qD) &= \delta \mathcal{D} \\ &= 2\xi'\mathcal{D} + \mathcal{D}'\xi - \frac{1}{2}\xi''' \\ &= 2\xi'qD + qD'\xi - \frac{1}{2}\xi''' \\ &= q\left( 2\xi'D + D'\xi - \frac{1}{2q}\xi''' \right) \end{split} \end{align}
or equivalently
\begin{equation} \label{eq:Diffeomorphism Field as Projective Gauge Field 2} \delta D = 2\xi'D + D'\xi - \frac{1}{2q}\xi''' \;. \end{equation}
Choosing $q=\frac{1}{2c}$,  we see a correspondence between the one-dimensional Thomas projective connection and the coadjoint element in Eq.~(\ref{coadjointaction}). 
  This improves the argument made in \cite{Brensinger:2017gtb}.
\subsection{Correspondence through two-cocycles}
The covariant metric allows us to improve upon another correspondence between the projective connection and coadjoint elements discussed in \cite{Brensinger:2017gtb}. We consider a projective 2-cocycle on \(\mathcal{N} \) for a path $C$ as
\begin{align}
<\xi,\eta>_{(\zeta)} = q \int_{C(\zeta)}& \xi^\a ({\tilde \nabla}_\a G^{\rho \n} {\tilde \nabla}_\rho {\tilde \nabla}_\n \eta^\b\ G_{\b \m} )\zeta^\m d\s\nonumber\\
 &-(\xi \leftrightarrow \eta)\;,
\end{align}  where $\s$ parameterizes the path.  The vector $\zeta^\mu \equiv\frac{d x^\m}{d\s}$ defines the path $C$. Here, the coordinates on  \(\mathcal{N} \) are $x^\a=(x,\l). $ We choose the vector fields as $\xi^\b = (\xi^b,-\l \xi^a g_a )$ and $\eta^\b = (\eta^b,-\l \eta^a g_a )$.  Consider a path given by a fixed value  $\l=\l_{0}$ along the vector  $\zeta_{\l_0}^{\mu} = ( \frac{dx}{d\s},0)$.  The metric used to construct the projective  Laplacian  is the one-dimensional version of Eq.~(\ref{eq:Big Metric Inverse}). Setting the metric to a constant $ g_{11}$ and the components of the vector fields to $\xi_1$ and $\eta_1$, respectively
 and keeping in mind that $\Pi^a_{\,\,\,bc}=0$ in one dimension, one finds that 
\begin{align}
<\xi,\eta>_{(\zeta_{\l_0})} \ =&\ q \int \xi_1 \left(2 \mathcal{D}_{11}- g_{11} \frac{1}{\l_0^2 } \right)\eta'_1 dx 
\nonumber\\
&+ q \int \xi_1 \eta_1^{'''}dx -(\xi \leftrightarrow \eta)\;.
\label{projective2cocycle}
\end{align}
 Comparing this to Eq.~(\ref{2cocycle2}), we make the observation that the projective connection and the coadjoint element $(B,q)$ are in correspondence through    
\begin{equation}
2\, q\,  \mathcal{D}_{11}-  \frac{q}{\l_0^2 } =  B\;,
\end{equation}    
 which recovers Eq.~(\ref{2cocycle2}) for \(q=\frac{c}{2 \pi}\). 
\subsection{Correspondence through gauge invariant action}
Using the action in \cite{Brensinger:2017gtb}, we  write the invariant projective Einstein-Hilbert terms as 
\begin{align} \label{eq:TwoD Projective EH Action} \begin{split} S_{\text{PEH}} &= \int d^2x \; d\l \sqrt{|G|} \cK_{\a\b} G^{\a \b}    = \int d^2x\; d\lambda \sqrt{|G|}   \cK   \\ &= \lambda_0 \left[\int d\lambda \ \frac{1}{\lambda} \ \right] \int \ d^2x \sqrt{|g|} \ (R+g^{ab}(2 \mathcal{P}_{ba}-\mathcal{P}_{ab}))  \\ &= \beta \left(S_{\text{EH}} + \int d^2x \sqrt{|g|} \ g^{ab}(2\mathcal{P}_{ba}-\mathcal{P}_{ab})  \right)\;, \end{split} \end{align} where we have used the projective Schouten tensor to write this in terms of the Riemann scalar curvature for familiarity. In two-dimensions, the Einstein-Hilbert term is the Gauss-Bonnet topological invariant.  The Polyakov metric has constant volume and $\cD_{ab}$ and $\cP_{ab}$ are equivalent.  Evaluating this on the Polyakov metric in two dimensions gives the coupling to the coadjoint element
\[ S_{\text{Polyakov Coupling}} = \int d^2\theta \cP_{++} h_{--}\;.\]
Again, the importance of this is to show dimensional universality of the interaction term in the Polyakov action as \( \sqrt{|G|} \ \cK\) has meaning in any dimension.  Thus, $\cD_{ab}$ is  to the Virasoro algebra of one-dimensional centrally extended vector fields as the Yang-Mills gauge field $A_a$ is to affine Lie algebras in one-dimension.  Furthermore, the projective curvature \( \cK^\a{}_{\m\n\b} \) can be used to build dynamical theories for \( \cD_{ab}\) just as the gauge curvature \( F_{ab}\) can provide dynamics for the gauge fields related to external gauge symmetries.    
\section{ Spinor Fields on  \texorpdfstring{$\mathcal{N}$}{N}} \label{chapter:Projective Fermions}

To this point, we have discussed the representation theory for the Thomas-Whitehead connection as related to tensors. Now we examine the relation among projective connections, spinors, and their associated Dirac equation. We will focus on spin $\frac{1}{2}$ spinors throughout. 
\subsection{The spin covariant derivative }

 To construct the spin connection for the generalized metric $G_{\a \b}$ we will  need the frame fields that make contact with the Minkowski space metric on the Thomas Cone.  There are several types of indices involved. First, there is a distinction between spacetime indices and the extra \(\lambda\) coordinate on \(\mathcal{N}\). Second, there is a distinction between curved indices and flat indices. To make calculations clear, we will adopt the following conventions for indices:
\begin{align*} 
\begin{split} &\text{(1)} \ \text{Greek indices } \mu, \nu - \text{curved coordinates on $\mathcal N$ }  \\ &\text{(2)} \ \text{Latin indices } m, n - \text{curved coordinates on $\mathcal M$ }  \\ &\text{(3)} \ \text{Underlined indices } \underline a, \underline \a - \text{flat coordinates on $\mathcal M$, $\mathcal N$ } \\ &\text{(4)} \ \lambda - \text{volume bundle coordinate} \\ &\text{(5)} \ \underline5 - \text{flattened volume bundle coordinate} \end{split} \end{align*}
We use the number \(\underline 5\) to represent the extra coordinate in flat space, due to the case of four-dimensional spacetime, where the gamma matrices are commonly labeled \(0,1,2,3,5\) for historical reasons.
For the metric $g_{m n}$ on $\mathcal{M}$ ,    \(e^{m}_{\ \underline  a}\) (with inverse \(e^{\underline  a}_{\ m}\)) are the associated frame fields  satisfying
 \begin{align} \begin{split} \label{eq:Frame Fields on Metric} g_{m n} &= e^{\underline  a}_{\ m}e^{\underline b}_{\ n} \eta_{\underline a\underline b} \\ \eta_{{\underline a} \underline b} &= e^{m}_{\ \underline a} e^{n}_{\ \underline b} g_{m n} \;. \end{split} \end{align}
Similarly, the frame fields denoted  \(\tilde{e}^{\mu}_{\ \underline \a}\) will be associated with the metric \(G_{\mu \nu}\) on \(\mathcal{N}\), and the indices range over all dimensions, including \(\lambda\) for the curved coordinates on \(\mathcal{N}\)
 \begin{align} \begin{split} \label{eq:Frame Fields on GMetric} G_{\a \b} &= {\tilde e}^{\underline \m}_{\ \a} \  {\tilde e}^{\underline \n}_{\ \b} \ \tilde \eta_{\underline\m \underline\n} \\ \tilde \eta_{\underline\a \underline\b} &= \tilde e^{\mu}_{\ \underline\a} \tilde e^{\nu}_{\ \underline\b} G_{\mu \nu} \;. \end{split} \end{align}
We may also use the frame fields to write the components of the Dirac matrices in curved spacetime coordinates
\begin{equation} \label{eq:Curved Dirac Matrices} \tilde \gamma^{\mu} = \tilde e^{\mu}_{\ \underline \a} \tilde \gamma^{\underline\a}\;. \end{equation}For the metric \(G_{\mu \nu}\)  given by Eq.~(\ref{eq:Big Metric}), the frame fields are listed as follows: \begin{align} \label{eq:Frame Fields on VM} \begin{split} \tilde{e}^{\,m}_{\,\ \underline a} &= e^{m}_{\ \underline a} \\ \tilde{e}^{\,m}_{\ \underline5} &= 0 \\ \tilde{e}^{\lambda}_{\ \underline a} &= -\lambda e^{m}_{\ \underline a}g_{m} \\ \,\,\,\tilde{e}^{\lambda}_{\ \underline5} &= \frac{\lambda}{\lambda_0} \;. \end{split} \end{align}
 The inverse frame field components are then given by:
\begin{align} \label{eq:Inverse Frame Fields on VM} \begin{split} \tilde{e}^{\,\underline a}_{\ m} &= e^{\underline a}_{\ m} \\  \tilde{e}^{\,\underline 5}_{\ m} &= \lambda_0 g_{m} \\ \tilde{e}^{\,\underline a}_{\ \lambda} &= 0\,\,\ \\ \tilde{e}^{\,\underline 5}_{\ \lambda} &= \frac{\lambda_0}{\lambda} \;. \end{split} \end{align}
Let \(\tilde\Gamma^{\mu}_{\ \nu \rho}\) be the components of the TW  connection, and call \(\mathring{\nabla}_{\mu}\) the corresponding covariant derivative operator that acts only on the curved indices as opposed to flat indices.  Define
\begin{equation} \label{eq:Spin Connection Components} \tilde\omega^{\mu}_{\ \underline\a \nu} = \mathring{\nabla}_{\nu}\tilde e^{\mu}_{\ \underline\a} = \partial_{\nu}\tilde e^{\mu}_{\ \underline\a} + \tilde \Gamma^{\mu}_{\ \rho \nu} \tilde e^{\rho}_{\ \underline\a} \;. \end{equation}
We use the geometric object \(\tilde \omega\) of  Eq.~(\ref{eq:Spin Connection Components}) to define a new \textit{spin covariant derivative}
\begin{align} \begin{split} \label{eq:Spin Covariant Derivative} \tilde D_{\mu} V^{\underline\a} &= \partial_{\mu} V^{\underline\a} + \tilde \omega^{\underline\a}_{\ \underline\b \mu}V^{\underline\b} \\ \tilde D_{\mu} V_{\underline\a} &= \partial_{\mu} V_{\underline\a} - \tilde \omega^{\underline\b}_{\  {\underline\a}  \mu}V_{\underline\b}\;,  \end{split} \end{align}
which recognizes tensorial objects, such as the vector \(V^{\underline\a}\), written in flat coordinates. 
We can take the full covariant derivative of a geometric object with curved \textit{and} flat spacetime indices by using the ordinary connection coefficients \(\tilde\Gamma^{\mu}_{\ \nu \rho}\) for curved indices and the spin connection coefficients \(\tilde \omega^{\underline\mu}_{\ \underline\a \nu}\) for flat indices. From now on, we will denote this full covariant derivative operator by \(\tilde \nabla_{\mu}\). By construction, the frame fields are covariantly constant, satisfying
\begin{equation} \label{eq:Frame Field Spin Covariant Derivative} \tilde \nabla_{\nu} \tilde e^{\mu}_{\ \underline\a} = 0 = \tilde \nabla_{\nu} \tilde e^{\underline\a}_{\ \mu}\;. \end{equation}
Then for any vector \(V^{\mu}\), we have
\begin{equation} \label{eq:Pulling Frame Field out of Derivative} \tilde \nabla_{\nu}V^{\underline\a} = \tilde e^{\underline\a}_{\ \mu}\tilde \nabla_{\nu}V^{\mu}\;, \end{equation}
so that the frame fields can be used to change indices without having to introduce an extra derivative term. 

With the frame fields on hand, we can calculate the coefficients of the TW spin connection using Eq.~(\ref{eq:Spin Connection Components}). Recall the TW connection coefficients originally presented in Eq.~(\ref{e:Gammatilde}):
\begin{align} \label{eq:TW Coefficients Again} \begin{split} &\tilde{\Gamma}^a_{\ bc} = \Pi^{a}_{\ bc} \\ &\tilde{\Gamma}^{\lambda}_{\ bc} = \lambda \mathcal{D}_{bc} \\ &\tilde{\Gamma}^a_{\ \lambda b} = \tilde{\Gamma}^{a}_{\ b \lambda} = \frac{1}{\lambda}\delta^{a}_{\ b} \end{split}
\end{align}
We simply need to plug these coefficients and the frame fields into Eq.~(\ref{eq:Spin Connection Components}) to get the TW spin coefficients that we desire. For example, if \(\underline a,\underline b \neq \underline 5\) and \(\mu \neq \lambda\) (which aligns with our chosen index conventions), we have
\begin{align} \label{eq:Calculation of TW Spin Coefficient} \begin{split} \tilde{\omega}_{\underline a \underline b m} &= \left( \partial_{m}\tilde{e}^{n}_{\ \underline b} + \tilde{\Gamma}^{n}_{\ \rho m} \tilde{e}^{\rho}_{\ \underline b} \right) \tilde{e}^{\underline c}_{\ n} \eta_{\underline a \underline c} \\
& \ \ \ \ + \left( \partial_{m}\tilde{e}^{\lambda}_{\ \underline b} + \tilde{\Gamma}^{\lambda}_{\ \rho m} \tilde{e}^{\rho}_{\ \underline b} \right) \tilde{e}^{\underline c}_{\ \lambda} \eta_{\underline a \underline c} \\ & \ \ \ \ + \left( \partial_{m}\tilde{e}^{n}_{\ \underline b} + \tilde{\Gamma}^{n}_{\ \rho m} \tilde{e}^{\rho}_{\ \underline b} \right) \tilde{e}^{\underline 5}_{\ n} \eta_{\underline a \underline 5} \\ & \ \ \ \ + \left( \partial_{m}\tilde{e}^{\lambda}_{\ \underline b} + \tilde{\Gamma}^{\lambda}_{\ r m} \tilde{e}^{r}_{\ \underline b} \right) \tilde{e}^{\underline 5}_{\ \lambda} \eta_{\underline a \underline 5}\;, \end{split} \end{align}
and since \(\eta_{\underline a \underline 5} = 0\) and \(\tilde{e}^{\underline c}_{\ \lambda} = 0\), this reduces to
\begin{align} \label{eq:Calculation of TW Spin Coefficient 2} \begin{split} \tilde{\omega}_{\underline a\underline b m} &= \left( \partial_{m}\tilde{e}^{\nu}_{\ \underline b} + \tilde{\Gamma}^{\nu}_{\ \rho m} \tilde{e}^{\rho}_{\ \underline b} \right) \tilde{e}^{\underline c}_{\ \nu} \eta_{\underline a \underline c} \\ &= \left( \partial_{m}e^{\nu}_{\ \underline b} + \Pi^{\nu}_{\ \rho m} e^{\rho}_{\ \underline b} \right) e^{\underline c}_{\ \nu} \eta_{\underline a \underline c} \\ &= \left( \partial_{m}e^{n}_{\ \underline b} + \Gamma^{n}_{\ r m} e^{r}_{\ \underline b} + \delta^{n}_{\ r}\a_{m}e^{r}_{\ \underline b} + \delta^{n}_{\ m}\a_{r}e^{r}_{\ \underline b} \right) e^{\underline c}_{\ n} \eta_{\underline a \underline c} \\ &= \omega_{\underline a \underline b m} + \a_{m} \eta_{\underline a \underline b} + e^{\rho}_{\ \underline b}e^{\underline c}_{\ m} \a_{\rho}\eta_{\underline a \underline c}\;, \end{split} \end{align}
where \(\omega_{\underline a\underline b m}\) (without a tilde) is the coefficient of the spin connection for the underlying spacetime connection \(\Gamma^{n}_{\ p m}\). Thus, we see how the TW spin connection coefficients \(\tilde{\omega}_{\underline a \underline b m}\) are offset from the spacetime spin connection coefficients \(\omega_{\underline a\underline b m}\) for \(\underline a,\underline b \neq \underline 5\). Below, we present the full list of independent TW spin connection coefficients:
\begin{align} \label{eq:List of TW Spin Connection Coefficients} \begin{split} \tilde{\omega}_{\underline a \underline b m} &= \omega_{\underline a\underline b m} + \a_{m} \eta_{\underline a\underline b} + e^{\rho}_{\ \underline b}e^{\underline c}_{\ m} \a_{\rho} \eta_{\underline a\underline c} \\ \tilde{\omega}_{\underline a\underline b \lambda} &= \frac{1}{\lambda}\eta_{\underline a\underline b} \\ \tilde{\omega}_{\underline a\underline 5 m} &= 0 \\ \tilde{\omega}_{\underline 5\underline b m} &= -\lambda_0 e^{
p}_{\ \underline b} \left( \mathcal{D}_{p m} - \partial_{m}g_{p} + \Gamma^{n}_{\ p m} g_{n} + \a_{m}g_{p} + \a_{p}g_{m} \right) \\ \tilde{\omega}_{\underline a\underline 5 \lambda} &= \tilde{\omega}_{\underline 5\underline b \lambda} = \tilde{\omega}_{\underline 5\underline 5 m} = 0 \\ \tilde{\omega}_{\underline 5\underline 5 \lambda} &= - \frac{1}{\lambda}\;. \end{split} \end{align}
In Eqs.~(\ref{eq:Calculation of TW Spin Coefficient 2}) and (\ref{eq:List of TW Spin Connection Coefficients}), we have explicitly written out $\Pi^{\nu}_{\ \rho \mu} $ in terms of a member of the equivalence class, $\Gamma^{a}_{\ bc} $ and its trace $\a_a$.  This allows us to see the relationship to the spin connection on $\mathcal M$. It is clear that $\tilde \o_{\underline a \underline b \rho}$ is not anti-symmetric in $\underline a$ and $\underline b$. 
\subsection{The  spinor connection  }
Let \( \Psi(x)\) and \(\bar \Psi(x)\) be a spinor  field and its Pauli adjoint, respectively representing a fermion and its anti-partner on the  manifold \(\mathcal{\mathcal{N}}\).  Then the covariant derivative acting on the spinor is  
\begin{equation} \label{eq:Spinor Covariant Derivative} \tilde \nabla_{\mu} = \partial_{\mu} + \tilde \O_{\mu}\;, \end{equation}where
\begin{equation} \label{eq:Spinor Connection Coefficients} \tilde \O_{\mu} = \frac{1}{4}\tilde \omega_{\underline \a \underline \b \mu} \tilde\gamma^{\underline \a} \tilde\gamma^{\underline \b} \end{equation}
and 
\begin{equation} \label{eq:Explicit Spinor Covariant Derivative} \tilde \nabla_{\mu}  \Psi = \partial_{\mu} \Psi + \tilde \O_{\mu} \Psi = \partial_{\mu} \Psi + \frac{1}{4} \tilde \omega_{\underline \a \underline \b \mu} \tilde \gamma^{\underline \a} \tilde\gamma^{\underline \b} \Psi\;.  \end{equation}
Similarly,
\begin{equation}
\nabla_{\mu}  \bar \Psi  = \partial_{\mu} \bar \Psi - \bar\Psi \tilde \O_{\mu}  = \partial_{\mu} \bar \Psi - \frac{1}{4}  \tilde \omega_{\underline \a \underline \b \mu} \bar \Psi \tilde \gamma^{\underline \a} \tilde\gamma^{\underline \b} \;.
\end{equation}The spin connection in  Eqs.~(\ref{eq:Spinor Connection Coefficients}) and (\ref{eq:Explicit Spinor Covariant Derivative}) have in general both symmetric and antisymmetric components in their flat indices \(a,b\). This is because the connection \(\tilde \Gamma^{\mu}_{\ \nu \rho}\) on $\mathcal N$ is not a metric compatible connection, since \( \tilde \nabla\) cannot be made metric compatible. The enveloping algebra of the gamma matrices is thus 
\begin{equation}
\tilde \gamma^{\underline \a} \tilde\gamma^{\underline \b} =-i  \s^{\underline\a\underline \b} + \tilde \eta^{\underline\a \underline\b}I_{4}\;,
\end{equation}
where the Sigma matrices generate the local $SO(4,1)$ Lorentz algebra on the Thomas cone, i.e.
\begin{equation}
[\s^{\underline \a \underline\b},\s^{\underline\m \underline\n}] = - 2 i (\eta^{\underline\a \underline\m}\s^{\underline \b \underline\n} + \eta^{\underline\b \underline\n}\s^{\underline\a \underline\m}-\eta^{\underline\a \underline\n}\s^{\underline\b \underline\m}-\eta^{\underline\b \underline\m}\s^{\underline\a \underline\n})\;.
\end{equation}
The $\tilde \omega_{[\underline \a \underline\b] \mu}$ therefore correspond to gauge fields for the local Lorentz transformation, while the $\tilde \omega_{(\underline\a  \underline\b) \mu}$ generate a translation on the fermions to their tensor densities. Let us write \( \O_\mu = \O^S_\mu + \O^A_\mu \), such that the symmetric component is
\begin{equation}
\tilde\O^S_{\rho} =\tilde \omega_{(\underline \a \underline\b) \rho}\tilde \eta^{\underline\a \underline\b} =(\rm{d}+1)\tilde\o_\rho \label{densitycurrent}
 \end{equation} and the expected SO(4,1) connection is 
\begin{equation} \tilde\O^A_{\rho} = - i\,\tilde \omega_{[\underline\a \underline\b] \rho}\tilde \,\s^{\underline \a \underline\b}\;.
\end{equation}
 In Eq.~(\ref{densitycurrent}), the space-time component of this Abelian connection is $\G^a_{\,\,\,a c}\equiv \G_c$.  In differential geometry, such a term appears in the presence of \emph{weighted spinors } \cite{Ogievetsky:1965ii}  that transform relative to an unweighted spinor $\phi$ as \begin{equation}
\phi_{vw} = \det(g)^{\frac{(v I_{4}+ w \g^5  )}{2}}\phi 
\end{equation}   
in four-dimensions. The spinor $\psi_{vw}$ is said to have weight $(v I_{4} + w \g^5)$.  For these weighted spinors, the spin connection is augmented to be\cite{Ogievetsky:1965ii}
\begin{equation}
\O_m \rightarrow \O_m + (v I_{4}+ w \g^5)\G_m\;.
\end{equation} 
We use this to define spinor representations (\( \frac{1}{2} \) integer spin) on the Thomas Cone. First, we remark that on the Thomas Cone $\g^5$ is an invariant tensor since \[\l_0 \g^5=\Upsilon^{\underline\a_1}\epsilon_{\underline\a_1 \cdots  \underline\a_5}\g^{\underline\a_2}\cdots\g^{\underline\a_5}, \]  where $\Upsilon^{\underline \n}= \tilde e^{\underline \n}_{\,\,\,\,\,\m}\Upsilon^{\m}.$ Since $\l \rightarrow \l |J|^{\frac{-1}{\rm{d}+1}}$, we can expect  weighted spinor representations on the Thomas Cone to be
 \be\Psi(x^a,\l)= \left(\frac{\l}{\l_0}\right)^{\frac{(\rm{d}+1)}{2}(v I_{4}+ w \g^5)}\phi(x^a) \label{ThomasSpinor} \ee and
   \be\bar \Psi(x^a,\l)= \bar \phi(x^a)\left(\frac{\l}{\l_0}\right)^{-\frac{(\rm{d}+1)}{2}(v I_{4}+ w \g^5)}\;. \label{ThomasAntiSpinor}\ee
Note that \[ \left(\frac{\l}{\l_0}\right)^{M} = e^{(\log{\left(\frac{\l}{\l_0}\right)}) \,M} \] for a matrix $M$ and 
\be
\l\, \frac{\partial}{\partial_\l}  \left(\frac{\l}{\l_0}\right)^M = M \left(\frac{\l}{\l_0}\right)^{M}\;.
\ee 
\subsection{The gauge invariant TW Dirac action}

Eq.~(\ref{eq:Explicit Spinor Covariant Derivative}) is the general expression for the covariant derivative of a spinor field. For the TW connection, we will use the spin connection coefficients given by Eq.~(\ref{eq:List of TW Spin Connection Coefficients}) and decompose the connection into its chiral and non-chiral parts.  This will illuminate the nature of the TW spinor connection on $\mathcal M$. 
With this,   $ \slashed \nabla$ becomes 
\begin{align} \label{eq:TW Spinor Covariant Derivative} \begin{split} \tilde{\slashed{\nabla}} \Psi = \tilde{\gamma}^\a \tilde{\nabla}_\a \Psi &= \tilde{\gamma}^{m}\partial_{m}\Psi + \tilde{\gamma}^{\lambda}\partial_{\lambda}\Psi \\ & \hspace{.5cm} + \frac{1}{4} \Big( \tilde{\gamma}^{m}\tilde{\omega}_{\underline a \underline b m}\tilde{\gamma}^{\underline a}\tilde{\gamma}^{\underline b} + \tilde{\gamma}^{\lambda}\tilde{\omega}_{\underline a \underline b \lambda}\tilde{\gamma}^{\underline a}\tilde{\gamma}^{\underline b} \\ & \quad \quad  \; \;+ \tilde{\gamma}^{m}\tilde{\omega}_{5 \underline b m}\tilde{\gamma}^5\tilde{\gamma}^{\underline b} + \tilde{\gamma}^{\lambda}\tilde{\omega}_{55 \lambda}\tilde{\gamma}^5\tilde{\gamma}^5 \Big) \Psi\;. \end{split} \end{align}
And for $\bar \Psi,$ 
\begin{align} \label{eq:TW Anti Spinor Covariant Derivative} \begin{split} \tilde{\slashed{\nabla}}\bar \Psi =  (\tilde{\nabla}_\a \bar \Psi) \tilde{\gamma}^\a &= \partial_{m}\bar \Psi\tilde{\gamma}^{m} + \partial_{\lambda}\bar \Psi\tilde{\gamma}^{\lambda} \\ & \hspace{.5cm} - \frac{1}{4}\bar \Psi \Big( \tilde{\omega}_{\underline a \underline b m}\tilde{\gamma}^{\underline a}\tilde{\gamma}^{\underline b} \tilde{\gamma}^{m}+ \tilde{\omega}_{\underline a \underline b \lambda}\tilde{\gamma}^{\underline a}\tilde{\gamma}^{\underline b}\tilde{\gamma}^{\lambda} \\ & \hspace{1.0cm} + \tilde{\omega}_{5 \underline b m}\tilde{\gamma}^5\tilde{\gamma}^{\underline b}\tilde{\gamma}^{m} + \tilde{\omega}_{55 \lambda}\tilde{\gamma}^5\tilde{\gamma}^5 \tilde{\gamma}^{\lambda}\Big) \;. \end{split} \end{align}

Evaluating these with the  coefficients from Eq.~(\ref{eq:List of TW Spin Connection Coefficients}) yields
\begin{align} \label{eq:TW Spinor Covariant Derivative 2}  \tilde{\slashed{\nabla}} \Psi = &\slashed{\nabla} \Psi - i\frac{\lambda}{\lambda_0}\gamma^{5}\left(\partial_{\lambda} \Psi\right) - \lambda g_{m} \gamma^{m} \left(\partial_{\lambda}\Psi\right)\nonumber \\
 &+ \frac{1}{4} \Big[ {\rm{d}}(\alpha_{m}-g_{m})+{\rm{d}}\alpha_{m}-g_{m} \Big] \gamma^{m}\Psi \nonumber\\
 &- \frac{1}{4} i \bigg[ \lambda_{0} \big( \mathcal{D}_{r m} - \partial_{m}g_{r} + \Gamma^{n}_{\ r m}g_{n} + \alpha_{m}g_{r} + g_{m}\alpha_{r} \big) g^{m r}\nonumber\\
 &+ \frac{1}{\lambda_0}(\rm{d}+1) \bigg] \gamma^5 \Psi\;,   \end{align} whilst
\begin{align} \label{eq:TW Anti Spinor Covariant Derivative 2} 
\tilde{\slashed{\nabla}} \bar \Psi = &\slashed{\nabla} \bar \Psi + i\frac{\lambda}{\lambda_0}\left(\partial_{\lambda} \bar\Psi\right)\gamma^{5} + \lambda g_{m} \left(\partial_{\lambda}\bar\Psi\right)\gamma^{m} \nonumber \\ 
&- \frac{1}{4} \Big[ {\rm{d}}(\alpha_{m}-g_{m})+{\rm{d}}\alpha_{m}-g_{m} \Big] \bar\Psi\gamma^{m} \nonumber\\ 
&+ \frac{1}{4} i \bigg[ \lambda_{0} \big( \mathcal{D}_{r m} - \partial_{m}g_{r} + \Gamma^{n}_{\ r m}g_{n} + \alpha_{m}g_{r} + g_{m}\alpha_{r} \big) g^{m r}\nonumber\\
&+ \frac{1}{\lambda_0}({\rm{d}}+1) \bigg]\bar \Psi\gamma^5 \;. 
\end{align} 
Here,  \(\nabla_{m}\) (without a tilde) is the spinor covariant derivative operator associated with the space-time connection \(\Gamma^{m}_{\,\,\ n r}\). Using this decomposition, we write Eq.~(\ref{eq:TW Spinor Covariant Derivative 2}) as 
\begin{align} \label{eq:TW Spinor Covariant Derivative Decomposed}  
\tilde{\slashed{\nabla}} \Psi = &\slashed{\nabla} \Psi - i\frac{\lambda}{\lambda_0}\gamma^{5}\left(\partial_{\lambda} \Psi\right) - \lambda g_{m} \gamma^{m} \left(\partial_{\lambda}\Psi\right)\nonumber\\
 &+ B_m \gamma^{m}\Psi -i \frac{1}{4}\Xi\ \g^5\ \Psi\;, 
\end{align}
where we've defined  $B_m$ and $\Xi$ as
\begin{align}
B_m &\equiv  \frac{1}{4} \Big( {\rm{d}}(\alpha_{m}-g_{m})+{\rm{d}}\alpha_{m}-g_{m} \Big)\\
\Xi &\equiv   \lambda_{0} \left( \mathcal{D}_{r m} - \partial_{m}g_{r} + \Gamma^{n}_{\ r m}g_{n} + \alpha_{m}g_{r} + g_{m}\alpha_{r} \right) g^{m r}\nonumber\\
 &\quad \quad+ \frac{1}{\lambda_0}({\rm{d}}+1) \;.
\end{align}    

The \textit{TW Dirac Lagrangian Density} that yields the Dirac equation 
\begin{equation}
i\tilde{\slashed{\nabla}} \Psi - (M+i M_\chi \g^5) \Psi =0,
\end{equation}
for a mass $M$ and a chiral mass $M_\chi$, may be written explicitly in covariant  and self conjugate form as
\begin{align} \label{eq:TW Dirac Lagrangian Hermitean} 
\mathcal{L}_{\text{TWD}}  = &\frac{i}{2}\sqrt{|G|}G^{\m \n} \bigg(\bar{\Psi}\tilde\g_\m(\partial_\n + {\tilde\O}_\n )\Psi \nonumber\\
&\qquad \qquad \qquad\quad- (\partial_\n {\bar{\Psi}-\bar{\Psi} \tilde\O_\n )\tilde \g_\m \Psi}  \bigg)\nonumber\\
&- \sqrt{|G|}(M \bar{\Psi} \Psi + i M_\chi  \bar{\Psi} \g^5 \Psi)\nonumber\\
&- \frac{i}{2}\bar \Psi \tilde\nabla_\m (\sqrt{|G|}\tilde\g^\mu) \Psi\;.
\end{align}
The last term arises because the metric and covariant derivative operator are not compatible since \begin{equation}
 \tilde\nabla_\m \tilde\g^\mu= \partial_\m \tilde\g^\mu + \tilde\G^\mu_{\;\:\mu \a}\tilde\g^\a + [\tilde\O_\a , \tilde\g^\a]
\end{equation} does not vanish. The commutator term  is precisely where the field $\cD_{m n}$ resides.    We can rewrite this so that the field equations on $\Psi$  (or $\bar \Psi$) are explicit if we integrate by parts the derivative term on $\bar \Psi$.  Then 
\begin{align} 
 -\frac{i}{2}\sqrt{|G|}&G^{\m \n} (\partial_\n \bar{\Psi})\tilde \g_\m \Psi \nonumber\\
 &=-\partial_\n(\frac{i}{2}\sqrt{|G|}G^{\m \n}  \bar{\Psi}\tilde \g_\m \Psi)\nonumber\\
&\hspace{0.4cm}+ \bar{\Psi}\partial_\n(\frac{i}{2}\sqrt{|G|}G^{\m \n}  \tilde \g_\m \Psi)  \\
&=-\underbrace{{\partial_\l(\frac{i}{2}\sqrt{|G|}G^{\l \l}  \bar{\Psi}\tilde \g_\l \Psi)}}_{\text{total $\l$ Derivative}}\nonumber\\
 &\hspace{0.4cm}-\underbrace{\partial_a(\frac{i}{2}\sqrt{|G|}G^{a b}  \bar{\Psi}\tilde \g_b
 \Psi)}_{\text{total space-time derivative}} \nonumber \\
 &\hspace{0.4cm}+ \bar{\Psi}\partial_\n(\frac{i}{2}\sqrt{|G|}\tilde \g^\n \Psi) \;.\label{Total Derivative Term} 
\end{align} 
The total space-time derivative may be eliminated on the boundary.  However the total $\l$ derivative will in general be finite and could contribute to the field equations.  Let us examine this term more carefully.  One sees that 
\begin{align}
\partial_\l(\frac{i}{2}&\sqrt{|G|}G^{\l \l}  \bar{\Psi}\tilde \g_\l \Psi) =\partial_\l(\frac{i}{2}\sqrt{|G|}  \bar{\Psi}\tilde \g^\l \Psi)\\
&= -\partial_\l\left(\frac{i}{2}\frac{\l_0}{\l}\sqrt{|g|} \bar{\Psi} \frac{\lambda}{\lambda_0} \left( i \gamma^5 + \lambda_0g_{m}\gamma^{m}\right)\Psi\right)
\\
&= \frac{i}{2}\sqrt{|g|} \partial_\l\left(i\bar{\Psi}\g^5 \Psi+  \lambda_0g_{m}\bar{\Psi}\g^m \Psi  \right).\end{align}
From the spinor projective representations in Eqs.~(\ref{ThomasSpinor}) and (\ref{ThomasAntiSpinor}),  this will vanish when $w=0,$ eliminating any chiral density terms.  We also observe that the term, $g_m \bar \Psi \g^m \Psi$, would vanish if the coordinates were gauge fixed  so that $g_m=0$ (constant volume).  Had we used a constant volume metric, this condition would have gone unnoticed. This also guarantees that the action is a scalar.  The remaining term in Eq.~(\ref{Total Derivative Term}) leads to the last summand in the covariant Lagrangian density, Eq.~(\ref{eq:TW Dirac Lagrangian Hermitean}). 
With this, we write the Lagrangian  which realizes the Dirac equation on $\Psi$ as
\begin{align} \label{eq:TW Dirac Lagrangian Hermitean 2}
\mathcal{L}_{\text{TWD}}  = &\frac{\l_0}{\l}\sqrt{|g|} \left(i\bar{\Psi} \slashed{\nabla} \Psi  + \frac{\lambda}{\lambda_0}\bar{\Psi}\gamma^{5} \partial_{\lambda} \Psi - i\lambda g_{m} \bar{\Psi}\gamma^{m} \partial_{\lambda}\Psi \right)\nonumber \\ 
&+ \frac{\l_0}{\l}\sqrt{|g|}\left( iB_{m} \bar{\Psi}\gamma^{m}\Psi + \Xi   \bar{\Psi} \gamma^5 \Psi \right)\nonumber\\
&-\frac{\l_0}{\l} \sqrt{|g|}(M \bar{\Psi} \Psi + i M_\chi  \bar{\Psi} \g^5 \Psi)\;.
\end{align} 

Had we wished to add a Yang-Mills potential to the action, we would have a term 
\begin{align}
\mathcal{L}_{\text{YM}}&= \sqrt{|G|} \bar{\Psi}\tilde\g^\m {\tilde A}_\m \Psi\nonumber\\
&=\sqrt{|G|}\left(  -i\frac{1}{\l_0}\bar{\Psi} \g^5\Psi + \bar{\Psi}\g^a A_a \Psi  \right)\;,
\end{align}
where a chiral mass term $M_{\mathcal A} = \frac{1}{\l_0}$ is induced.  This follows since the corresponding projective one-form for the matrix valued potential is 
\begin{equation}
\tilde A_\m = (A_a +g_a {\bf 1}, \frac{1}{\l} {\bf 1})\;,
\end{equation}
where $\bf{1}$ is in the center of the algebra. Then, using Eq.~(\ref{eq:Extra Gamma Matrix}), we have the result
\begin{equation}
\tilde \g^\m{\tilde A_\m} = -i\frac{1}{\l_0} \g^5\, {\bf 1}+ \g^a A_a \;.
\end{equation} 

In the Lagrangian density, Eq.~(\ref{eq:TW Dirac Lagrangian Hermitean 2}), we have left terms with explicit  \(\lambda\) dependence of \(\Psi\). Following Eqs.~(\ref{ThomasSpinor}) and (\ref{ThomasAntiSpinor}), along with the requirement that the action be a scalar, we have
\begin{align} \label{eq:Spinor on VM} \begin{split} \Psi(x,\lambda) &= \left(\frac{\l}{\l_0}\right)^{\frac{v}{2}({\rm{d}}+1)}\phi(x^a) \\ \bar{\Psi}(x,\lambda) &= \left(\frac{\l}{\l_0}\right)^{-\frac{v}{2}({\rm{d}}+1)} \bar{\phi}(x)\;, \end{split} \end{align}
where \(v\) is the density weight which determines precisely how \(\Psi\) will transform under \(\lambda \to \lambda '\). In the TW Dirac Lagrangian \(\mathcal{L}_{\text{TWD}}\) of Eq.~(\ref{eq:TW Dirac Lagrangian Hermitean 2}), this representation of \(\Psi\) will only affect the terms
\begin{align} \label{eq:Lambda Derivative of Spinor} 
\frac{\lambda}{\lambda_0} \bar{\Psi} \gamma^5 \left(\partial_{\lambda}\Psi\right) -& i \lambda g_{m} \bar{\Psi} \gamma^{m} \left( \partial_{\lambda}\Psi \right)\nonumber\\
 &= \frac{v({\rm{d}}+1)}{2\lambda_0}  \bar{\Psi} \gamma^5 \Psi - i\frac{ v}{2}({\rm{d}}+1) g_{m}  \bar{\Psi} \gamma^{m} \Psi \;. 
\end{align}
so that the TW Dirac Lagrangian can be reduced to a Lagrangian on $\psi$ with $v$ a weight parameter
\begin{align} \label{eq:TW Dirac Lagrangian 3}
\mathcal{L}_{\text{TWD}} &= \frac{\l_0}{\l}\sqrt{|g|}( i \bar{\phi} \tilde{\slashed{\nabla}} \phi - M \bar{\phi} \phi)\nonumber \\ 
&= \frac{\l_0}{\l}\sqrt{|g|}\left( i \bar{\phi} \slashed{\nabla} \phi - M \bar{\phi} \phi \right)\nonumber\\ 
&\quad + \frac{\l_0}{\l}\sqrt{|g|}\bigg(\frac{1}{4} \Big[ {\rm{d}}(\alpha_{m}-g_{m})+{\rm{d}}\alpha_{m}-g_{m}\nonumber\\
&\quad \quad \quad \quad \quad \quad \quad- 2 g_{m} v({\rm{d}}+1) \Big] i \bar{\phi}\gamma^{m}\phi \nonumber \\ 
&\quad + \frac{1}{4} \Big[ \lambda_{0} \big( \mathcal{D}_{r m} - \partial_{m}g_{r} + \Gamma^{n}_{\ r m}g_{n} + \alpha_{m}g_{r} \nonumber\\
& \quad \quad \quad \quad \quad \quad + g_{m}\alpha_{r} \big) g^{m r} + \frac{1}{\lambda_0}({\rm{d}}+1)\nonumber\\
&\quad \quad \quad \quad \quad \quad+ 2 \frac{v({\rm{d}}+1)}{\lambda_0}  \Big] \bar{\phi} \gamma^5 \phi\bigg)\;.
\end{align}
A special choice of the weight $v=-\frac{1}{2}$ eliminates the induced chiral mass term (in the absence of gauge fields)  and also eliminates the metric density contribution in the coupling to $i \bar{\phi}\gamma^{m}\phi$.  

The only $\l$ dependence is in the overall coefficient $\frac{\l_0}{\l}\sqrt{|g|}$.   As we will discuss in Section  \ref{Gauge Invariant Action}, we may write $\ell \equiv \lambda/\lambda_0$ to be a dimensionless scale. By writing  $\sqrt{|G|}=\frac{1}{\ell}\sqrt{|g|}$, we have  
\[  \int d\l \;\frac{\l_0}{\l}  = \l_0 \int d\ell \;\frac{1}{\ell} =\l_0 \log(\ell_f/\ell_i) \; ,  \]
where $\ell_i$ and $\ell_f$ are original and final length scales.  
 With this, we can make  a field redefinition of the fermions $\phi $ and define  \(\psi = \phi \sqrt{\l_0 \log(\ell_f/\ell_i)} \  \) so that the fermions $\psi$ have the dimensions of four-dimensional fermions.  
The four-dimensional TW Dirac action  becomes
\begin{align} \label{eq:TW Dirac Action Renormalized}
&{S}_{\text{TWD}}= \int d^{\rm d}x \sqrt{|g|}\left( i \bar{\psi} \slashed{\nabla} \psi - M \bar{\psi} \psi \right)\\ 
 &+ \int d^{\rm d}x \sqrt{|g|}\bigg(\frac{1}{4} \Big[ {\rm{d}}(\alpha_{m}-g_{m})+{\rm{d}}\alpha_{m}-g_{m} \nonumber\\
 &\hspace{2.8cm}- 2 g_{m} v({\rm{d}}+1) \Big] i \bar{\psi}\gamma^{m}\psi\nonumber\\
& \hspace{0.2cm} + \frac{1}{4} \Big[ \lambda_{0} \left( \mathcal{D}_{r m} - \partial_{m}g_{r} + \Gamma^{n}_{\ r m}g_{n} + \alpha_{m}g_{r} + g_{m}\alpha_{r} \right) g^{m r}\nonumber \\ 
&\hspace{3.2cm} + \frac{1}{\lambda_0}({\rm{d}}+1) + 2 \frac{v({\rm{d}}+1)}{\lambda_0}   \Big] \bar{\psi} \gamma^5 \psi\bigg)\;. \nonumber 
\end{align}
We see that $\l_0$ still sets the chiral scale due to its presence in the last two summands of the action. 

 In the discussion following Eq.~(\ref{eq:VM Gamma Matrices in 4D}), we noted that we should expect a dynamical theory of \(\mathcal{D}_{mn}\) to be sensitive to chirality of fermions. This expectation is realized by the TW Dirac Lagrangian,  Eq.~(\ref{eq:TW Dirac Lagrangian Hermitean 2}), due to the presence of  \(\gamma^5\).  The theory is therefore \textit{chiral} in this sense. We remark that one can still eliminate $\rm d$ degrees of freedom by using a coordinate gauge choice.  For example, we could set  $g_a=0$ (constant volume gauge for the metric), $\a_a=0$ (constant volume for the connection) or even $g_a= \a_a$  (compatibility of condition) in Eq.~(\ref{eq:TW Dirac Action Renormalized}).  However, no gauge choice will eliminate the $\cD_{ab}$ fermion interaction. 
\section{ Gauge Invariant TW Action}\label{Gauge Invariant Action}

The TW Action was introduced in \cite{Brensinger:2017gtb} in order to give dynamics to the diffeomorphism field.  There, the correspondence with the coadjoint orbits of the Virasoro algebra was determined in the background of the gauged fixed  2D metric of Polyakov\cite{Polyakov:1987zb} that had constant volume.  Similarly in \cite{Brensinger:2019mnx}, the interest was to study the diffeomorphism field as a primeval source for dark energy in a Friedman-Lemaitre-Robertson-Walker background in constant volume coordinates.    As we have just seen in the Dirac action, writing the TW action in a gauge invariant form reveals physically interesting structure.  From \cite{Brensinger:2017gtb} the TW dynamical action is:
\begin{equation} \label{eq:Full Dynamical Action} S = S_{\text{PEH}} + S_{\text{PGB}} \end{equation}
where the projective Einstein-Hilbert action is 
\begin{equation} \label{eq:PEH Action} S_{\text{PEH}} = -\tfrac{1}{2\tkappa_0 \lambda_0} \int d\lambda\ d^{\rm{d}}x \sqrt{|G|} \cK^{a}_{\ bcd}(\delta^{c}_{\ a} g^{bd}) \end{equation}
and the projective Gauss-Bonnet action is
\begin{widetext}
\begin{equation} \label{eq:PGB Action} S_{\text{PGB}} =- \tfrac{\tJ_0 c}{\lambda_0} \int d\lambda \ d^{\rm{d}}x \sqrt{|G|} \ \left(\cK^{\alpha}_{\ \beta \gamma \rho}\cK_{\alpha}^{\ \beta \gamma \rho} - 4\cK_{\alpha \beta}\cK^{\alpha \beta} + \cK^2\right). 
\end{equation}
\end{widetext}
We remark that both terms are generalized Gauss-Bonnet terms and one could presumably continue adding generalized Gauss-Bonnet terms for higher interaction without compromising  causality in the metric field equations \cite{Lovelock:1971yv}.
Recall that the components of the TW curvature tensor \(\cK^{\alpha}_{\ \beta \gamma \rho}\) are given by
\begin{align} \label{eq:TW Curvature Components} \begin{split} &\cK^a_{\ bcd} = \mathcal{R}^a_{\ bcd} + \delta^a_{\ c} \mathcal{D}_{db} - \delta^a_{\ d}\mathcal{D}_{cb} \\ & \ \ \ \ \ \ \ = R^a_{\ bcd} + \delta^a_{\ c} \mathcal{P}_{db} - \delta^a_{\ d}\mathcal{P}_{cb} - \delta^a_{\ b}\mathcal{P}_{[cd]} \\ &\cK^{\lambda}_{\ bcd} = \lambda \left( \partial_{[c}\mathcal{D}_{d]b} + \Pi^a_{\ b[d} \mathcal{D}_{c]a} \right) \\ & \ \ \ \ \ \ \ = \lambda \bigg( \partial_{[c}\mathcal{P}_{d]b} + \Gamma^a_{\ b[d} \mathcal{P}_{c]a} + \alpha_{[d} \mathcal{P}_{c]b} \\
& \hspace{3.0cm}+ \alpha_b\mathcal{P}_{[cd]} - R^a_{\ bcd}\alpha_a \bigg) \\
&\breve{\mathcal{K}}\indices{_{cab}}\equiv \frac{1}{\lambda}\mathcal{K}\indices{^\lambda _{cab}}\;, 
\end{split} \end{align}
where again
\begin{align} \label{eq:Projective Curvature Tensor Components Again} \begin{split} \alpha_a &= -\frac{1}{{\rm{d}}+1}\Gamma^e_{\ ea} \\ \Pi^a_{\ bc} &= \Gamma^a_{\ bc} + \delta^a_{\ b}\alpha_c + \delta^a_{\ c}\alpha_b \\ \mathcal{R}^{a}_{\ bcd} &= \partial_c\Pi^a_{\ db} - \partial_d\Pi^a_{\ cb} + \Pi^a_{\ ce}\Pi^e_{\ db} - \Pi^a_{\ de}\Pi^e_{\ cb} \\ R^{a}_{\ bcd} &= \partial_c\Gamma^a_{\ db} - \partial_d\Gamma^a_{\ cb} + \Gamma^a_{\ ce}\Gamma^e_{\ db} - \Gamma^a_{\ de}\Gamma^e_{\ cb} \\ \mathcal{P}_{bc} &= \mathcal{D}_{bc} - \partial_b \alpha_c + \Gamma^e_{\ bc}\alpha_e + \alpha_b \alpha_c \end{split} \end{align}
for any affine connection \(\Gamma^a_{\ bc}\). The non-zero components of the TW Ricci tensor \(\cK_{\alpha \beta}\) are
\begin{align} \label{eq:TW Ricci Curvature Components} \begin{split} \cK_{bd} &= \mathcal{R}_{bd} + ({\rm{d}}-1)\mathcal{D}_{bd} \\ &= R_{bd} + \rd\mathcal{P}_{db} - \mathcal{P}_{bd}. \end{split} \end{align}
Then the projective Gauss-Bonnet action \(S_{\text{PGB}}\) may be decomposed as \[S_{\text{PGB}} = S_{\text{PGB1}} + S_{\text{PGB2}} + S_{\text{PGB3}}\;,\] where
\begin{widetext}
\begin{align} \begin{split} \label{eq:Parts of PGB Action} S_{\text{PGB1}} &= -\tfrac{\tJ_0 c}{\lambda_0} \int d^{\rm{d}}x \ d\lambda \sqrt{|G|} \ \cK^{a}_{\ bcd} \cK^{e}_{\ fgh} \left( \mathcal{B}_{ae}^{\ \ bfcgdh} - \lambda_0^{\ 2}g_{a}g_{e}g^{bf}g^{cg}g^{dh} \right) \\ S_{\text{PGB2}} &= \tJ_0 c \lambda_0 \int d^{\rm{d}}x \ d\lambda \sqrt{|G|} \ \breve{\mathcal{K}}_{bcd}\breve{\mathcal{K}}_{fgh} \left( g^{bf}g^{cg} g^{dh} \right) \\ S_{\text{PGB3}} &= 2\tJ_0 c \lambda_0 \int d^{\rm{d}}x \ d\lambda \sqrt{|G|} \  \cK^{a}_{\ bcd} \breve{\mathcal{K}}_{fgh} \left( g_{a} g^{bf}g^{cg}g^{dh} \right) \end{split} \end{align}
\end{widetext}
and we have defined the Gauss-Bonnet operator  as
\begin{align} \label{eq:Gauss-Bonnet Operator on N} \mathcal{G}_{\a \bar \a}^{\ \ \ \b \bar \b \g \bar \g \r \bar \r} = G_{\a \bar \a}G^{\b \bar \b}G^{\g \bar \g}G^{\r \bar \r} &- 4\delta^\g_{\ \a}\delta^{\bar \g}_{\ \bar \a}G^{\b \bar \b}G^{\r \bar \r}\nonumber\\
 &+ \delta^\g_{\ \a}\delta^{\bar \g}_{\ \bar \a}G^{\b \r}G^{\bar \b \bar \r} 
\end{align}
and for convenience, in terms of the metric on $\mathcal M$, 
\begin{align} \label{eq:Gauss-Bonnet Operator} \mathcal{B}_{a \bar a}^{\ \ \ b \bar b g \bar g r \bar r} = g_{a \bar a}g^{b \bar b}g^{g \bar g}g^{r \bar r} &- 4\delta^g_{\ a}\delta^{\bar g}_{\ \bar a}g^{b \bar b}g^{r \bar r} \nonumber\\
&+ \delta^g_{\ a}\delta^{\bar g}_{\ \bar a}g^{b r}g^{\bar b \bar r}\;. 
\end{align}
Finally, we can write the full dynamical action as
\begin{equation} \label{eq:Full Dynamical Action 2} S = S_{\text{PEH}} + S_{\text{PGB1}} + S_{\text{PGB2}} + S_{\text{PGB3}} \end{equation}
This form of the action is  convenient for computing field equations. The curvature components \(\cK^{a}_{\ bcd}\) and \(\breve{\mathcal{K}}_{bcd}\) carry all of the \(\Pi^a_{\ bc}\) and \(\mathcal{D}_{bc}\) (equivalently \(\Gamma^a_{\ bc}\) and \(\mathcal{P}_{bc}\)) dependence, while the metric tensor \(g_{ab}\) appears elsewhere in each part of the action, including in the Gauss-Bonnet operator \(\mathcal{B}\).
\\
\newpage
To illustrate explicit general coordinate invariance, it  is also possible to decompose the action as
\begin{widetext}
\begin{align} \label{eq:PGB Action Expanded} \begin{split} S &= \left( \int \frac{1}{\lambda} \ d\lambda \right) \Bigg[ -\tfrac{1}{2\tkappa_0} \int d^{\rm{d}}x \sqrt{|g|} \ \cK \\ & - \tJ_0 c \int d^{\rm{d}}x \sqrt{|g|} \left(\cK^{a}_{\ bcd}\cK_{a}^{\ bcd} - 4\cK_{ab}\cK^{ab} + \cK^2\right) \\ & +\tJ_0 c \lambda_0^{\ 2} \int d^{\rm{d}}x \sqrt{|g|} \underbrace{\left( g_a \cK^a_{\ bcd} + \breve{\mathcal{K}}_{bcd} \right)}_{\text{tensor}} \underbrace{\left( g_e \cK^e_{\ fgh} + \breve{\mathcal{K}}_{fgh} \right)}_{\text{tensor}} g^{bf}g^{cg}g^{dh} \Bigg]\;. \end{split} \end{align}
\end{widetext}
Eqs.~(\ref{eq:TW Curvature Components}) and (\ref{eq:TW Ricci Curvature Components}) demonstrate that \(\cK^a_{\ bcd}\), \(\cK_{ab}\), and \(\cK\) are tensors on the spacetime manifold \(\mathcal M\). Furthermore, we introduce $K_{bcd} \equiv \cK_{bcd}(g)$ as the following rank-three tensor on $\mathcal{M}$
\begin{align} \label{eq:Tensor in Projective Action} \begin{split}  K_{bcd} &\equiv  g_a\cK^a_{\ bcd} + \breve{\mathcal{K}}_{bcd}\\
 &=  \left( g_a-\alpha_a \right) R^a_{\ bcd} + \left(g_c-\alpha_c\right)\mathcal{P}_{db} - \left( g_d-\alpha_d \right)\mathcal{P}_{cb} \\ & \quad \; - \left( g_b-\alpha_b \right)\mathcal{P}_{[cd]} + \nabla_c \mathcal{P}_{db} - \nabla_d \mathcal{P}_{cb} \;, \end{split} \end{align}
where \(\nabla_a\) is the covariant derivative operator associated with the spacetime connection \(\Gamma^a_{\ bc}\). Since \(g_{a}\) and \(\alpha_a\) have the same coordinate transformation law, we see that $K_{bcd}$ is indeed a tensor on $\mathcal{M}$.  This demonstrates that the action is a  scalar as well as projectively invariant.

\label{Scaling} Owing to Eq.~(\ref{eq:Tensor in Projective Action}), all the $\l$ dependence appears as overall coefficients.  We will use the interpretation of the coupling constants as in \cite{Brensinger:2019mnx} to write them in terms of scale dependent quantities. Let $\ell \equiv \lambda/\lambda_0$ be a dimensionless scale.   Since  only $\int d\l\; \frac{1}{\l} $ appears in the overall coupling, we again write   $\sqrt{|G|}=\frac{1}{\ell}\sqrt{|g|}$.  Then by integrating over $\ell$, we can rewrite the action in terms of  coupling constants that have familiar interpretations
\begin{align}\label{e:flint}
        &\frac{1}{\tilde{\kappa}_0}\int_{\ell_i}^{\ell_f} d\ell\; \frac{1}{\ell} = \frac{\log(\ell_f/\ell_i)}{\tilde{\kappa}_0}\quad \Rightarrow \kappa_0 \equiv\frac{\tilde{\kappa_0}}{\log(\ell_f/\ell_i)}\\
        &\tilde{J}_0\int_{\ell_i}^{\ell_f}d\ell\; \frac{1}{\ell}=\tilde{J}_0\log(\ell_f/\ell_i)\quad \Rightarrow J_0 \equiv \tilde{J}_0 \log(\ell_f/\ell_i) \; .
\end{align}
Thus  a natural scaling of  the gravitational coupling constant $\kappa_0$ and angular momentum parameter $J_0$ occurs as we move from one length scale to another.
In this way, projective geometry has a potential renormalization group interpretation. This link is under further investigation. The characteristic projective length scale (inverse mass scale) is set by $\lambda_0$. With this, we can rewrite the 
TW action as 
\begin{align} \label{eq:PGB Action Rescaled} \begin{split} S &=   -\frac{1}{2\kappa_0} \int d^{\rm{d}}x \sqrt{|g|} \ \cK +   c J_0 \lambda_0^{\ 2} \int d^{\rm{d}}x \sqrt{|g|} K_{bcd}K^{bcd}  \\ &  -  c J_0 \int d^{\rm{d}}x \sqrt{|g|} \left(\cK^{a}_{\ bcd}\cK_{a}^{\ bcd} - 4\cK_{ab}\cK^{ab} + \cK^2\right) \end{split} \end{align}
where $K^{bcd}$ and $\cK_a{}^{bcd}$ have had the altitudes of their indices flipped via the metric and inverse metric on $\mathcal{M}$
\begin{align}
        K^{bcd} =& g^{bf} g^{cg} g^{dh}K_{fgh} ~~~,~~~        \cK_a{}^{bcd} = g_{am}g^{bf} g^{cg} g^{dh}\cK^m{}_{fgh} .
\end{align}

\section{The Covariant Field Equations} 
 In the spirit of Palatini \cite{palatini}, we will treat the metric tensor $g_{ab}$ and ${{\tilde\G}}^{\a}_{\,\,\b \g} $ as independent degrees of freedom.  This fits the framework of TW gravity, since the TW connection is to be thought of as a connection over the space of equivalence classes of connections and is not naturally tied to a particular metric.   The metric $G_{\m \n}$ serves only to maintain general coordinate invariance on $\mathcal N$, just as $\cD_{a b}$ exists in order to make the connection $\tilde \nabla_\m$ covariant. The covariant derivative is a projective invariant that  is constructed only from projectively invariant quantities such as  $\Pi^{a}_{\,\,\, b c}$ and $\l$. However, as one sees in Eq.~(\ref{e:Gammatilde}), the only degrees of freedom that are allowed to fluctuate are $\cD_{ab}$ and $\Pi^{a}_{\,\,\,bc}$. Therefore we will only need the field equations for $\Pi^{a}_{\,\,\,bc}$, $\cD_{ab}$, and $g_{ab}$. We note that $\l$ does not fluctuate and only sets the volume scale. 

We consider a total action of the form \begin{equation}
S_\text{total} = S + S_\text{matter},
\end{equation} 
where here $S$ is the TW action from Eq. (\ref{eq:Full Dynamical Action}) and $S_\text{matter}$ are contributions from other sources.  For example the Dirac action for each species of fermions will be in the form of Eq. (\ref{eq:TW Dirac Action Renormalized}) and could be accompanied by an appropriate gauge field action for  Yang-Mills fields.  Other matter contributions may also be considered.  In what follows, however,  we will derive the field equations from $S$ only with the understanding that the matter actions will also contribute non-trivially to these equations.  For the field equations of the metric, Section \ref{Subsec:Metric}, we will reinstate the matter contribution through the energy-momentum tensor, \( \Theta^\text{matter}_{pq}\).
 In \ref{Subsec:Palatini} we will demonstrate how one may use a \emph{Palatini} field, \(C^a_{\,\,\,bc}\), to utilize a metric compatible connection and recover the usual Einstein field equations with a divergence free energy-momentum tensor.  All the field equations in this section will be summarized in the Appendix.          
\subsection{Equations of motion for \texorpdfstring{\(\Pi^a_{\ bc}\)}{Pi}}

In order to simplify the computation of the field equations, we will use \(\mathcal{F}\) to denote an object with the correct valence to form a scalar with another given object. For example, we might write an expression such as \(\cK^a_{\ bcd} \mathcal{F}\), where we would understand that \(\mathcal{F}\) is an object with components \(\mathcal{F}_a^{\ bcd}\) such that \(\mathcal{F}\) forms a scalar upon tensor multiplication with \(\cK^a_{\ bcd}\).

With this, we compute the  field equations for \(\Pi^a_{\ bc}\) as:
\begin{align} \label{eq:Pi Variation 1} \begin{split} &S = \int \cK^a_{\ bcd}\mathcal{F} \\ &\implies \delta S = \int \left(\delta \cK^a_{\ bcd}\right)\mathcal{F} \\ &\qquad \qquad= \int \delta\left(\mathcal{R}^a_{\ bcd} + \delta^a_{\ c}\mathcal{D}_{db} - \delta^a_{\ d}\mathcal{D}_{cb}\right)\mathcal{F} \\ &= \int \delta \left(\partial_c\Pi^a_{\ db}-\partial_d\Pi^a_{\ cb}+\Pi^a_{\ ce}\Pi^e_{\ db}-\Pi^a_{\ de}\Pi^e_{\ cb}\right)\mathcal{F} \\ &= \int \bigg(-\delta^a_{\ l}\delta^m_{\ d}\delta^n_{\ b}\partial_c\mathcal{F} + \delta^a_{\ l}\delta^m_{\ c}\delta^n_{\ b}\partial_d\mathcal{F} + \big(\delta^a_{\ l}\delta^m_{\ c}\delta^n_{\ e}\Pi^e_{\ db} \\ & \qquad \ \ \ \  + \delta^e_{\ l}\delta^m_{\ d}\delta^n_{\ b}\Pi^a_{\ ce} - \delta^a_{\ l}\delta^m_{\ d}\delta^n_{\ e}\Pi^e_{\ cb} \\
&  \hspace{3.5cm}- \delta^e_{\ l}\delta^m_{\ c}\delta^n_{\ b}\Pi^a_{\ de}\big)\mathcal{F}\bigg) \delta \Pi^l_{\ mn} \end{split} \end{align}
and
\begin{align} \label{eq:Pi Variation 2} \begin{split} &S = \int \breve{\mathcal{K}}_{cab}\mathcal{F} \\ &\implies \delta S = \int \left(\delta \breve{\mathcal{K}}_{cab}\right)\mathcal{F} \\ &\hspace{1.4cm}= \int \delta \left(\partial_{[a}\mathcal{D}_{b]c} + \Pi^d_{\ c[b}\mathcal{D}_{a]d}\right)\mathcal{F} \\ &\hspace{1.4cm}= \int \left(\left(\delta^d_{\ l}\delta^m_{\ c}\delta^n_{\ b}\mathcal{D}_{ad} - \delta^d_{\ l}\delta^m_{\ c}\delta^n_{\ a}\mathcal{D}_{bd}\right)\mathcal{F}\right) \delta \Pi^l_{\ mn}\;. \end{split} \end{align}
These two variations lead to the full equations of motion for \(\Pi^a_{\ bc}\) that are associated with the appropriate object \(\mathcal{F}\) . We have
\begin{align} \label{eq:PEH Pi Variation} \begin{split} \delta S_{\text{PEH}} &= -\frac{1}{2\kappa_0} \int d^{\rm{d}}x \bigg[-\partial_{l}\left(\sqrt{|g|}g^{nm}\right) +\cancel {\delta^{m}_{\ l}\partial_e\left(\sqrt{|g|}g^{ne}\right)} \\ & \hspace{2.0cm}  + \sqrt{|g|} \big(\cancel{\delta^{m}_{\ l}\Pi^n_{\ db}g^{bd}} + \cancel{\Pi^{a}_{\ al}g^{nm}}\\
& \hspace{2.0cm} - \Pi^n_{\ l b}g^{b m} - \Pi^{m}_{\ d l}g^{n d}\big)\bigg] \delta \Pi^l_{\ mn}\;, \end{split} \end{align}
where the  striked-out terms vanish because  $\Pi^{a}_{\ b c}$ as well as its variation $\d\Pi^{a}_{\,\,\,bc}$ are traceless. The remaining contribution to the field equations would vanish if  $\Pi^{a}_{\ bc}$ were the traceless Levi-Civita connection of the metric $g_{ab}$, consistent with the original Palatini equations \cite{palatini}.
 The next contributions are 
 \begin{align} \label{eq:PGB1 Pi Variation} \begin{split} \delta S_{\text{PGB1}} &= -2J_{0}c \int d^{\rm{d}}x \left[\partial_{e}\left(\sqrt{|g|}\cK^a_{\ bcd}\left(\mathcal{B}_{la}^{\ \ nbmced}-\mathcal{B}_{la}^{\ \ nbecmd}\right.\right.\right. \\ & \hspace{0.5cm} \left.\left. + \lambda_0^{\ 2} g_l g_a g^{nb}g^{ec}g^{md} - \lambda_0^{\ 2} g_l g_a g^{nb}g^{mc}g^{ed} \right)\right) \\ & \hspace{0.5cm} \left. + \sqrt{|g|} \cK^a_{\ bcd} \left(\Pi^n_{\ e f}\left(\mathcal{B}_{la}^{\ \ fbmced} - \mathcal{B}_{la}^{\ \ fbecmd}\right.\right.\right. \\ & \hspace{0.5cm} \left. + \lambda_0^{\ 2} g_l g_a g^{fb}g^{ec}g^{md} - \lambda_0^{\ 2} g_l g_a g^{fb} g^{mc} g^{ed} \right) \\ & \hspace{0.5cm} \left.\left. + \Pi^{e}_{\ f l}\left(\mathcal{B}_{ea}^{\ \ nbfcmd} - \mathcal{B}_{ea}^{\ \ nbmcfd}\right.\right.\right. \\ & \hspace{0.5cm} \left.\left.\left. + \lambda_0^{\ 2} g_e g_a g^{nb}g^{mc}g^{fd} - \lambda_0^{\ 2} g_e g_a g^{nb}g^{fc}g^{md} \right)\right)\right] \delta \Pi^l_{\ mn} \end{split} \end{align}
\begin{align} \label{eq:PGB2 Pi Variation} \begin{split} \delta S_{\text{PGB2}} &= 4J_{0}c\lambda_0^{\ 2}  \int d^{\rm{d}}x \left[\sqrt{|g|} \breve{\mathcal{K}}_{fgh} \mathcal{D}_{cl} g^{mf}g^{cg}g^{nh} \right] \delta \Pi^l_{\ mn} \end{split} \end{align}
\begin{align} \label{eq:PGB3 Pi Variation} \begin{split} \delta S_{\text{PGB3}} &= 4J_{0}c\lambda_0^{\ 2}  \int d^{\rm{d}}x \left[ \partial_d \left( \sqrt{|g|} \breve{\mathcal{K}}_{fgh}g_l g^{nf}g^{mg}g^{dh} \right)\right. \\ &  \left. + \sqrt{|g|}\breve{\mathcal{K}}_{fgh} \left( g_l g^{bf}g^{mg}g^{dh}\Pi^{n}_{\ db} + g_a g^{nf}g^{cg}g^{mh}\Pi^{a}_{\ cl} \right) \right. \\ &  \left. + \sqrt{|g|} \cK^a_{\ bcd} g_a g^{bm}g^{cg}g^{dn}\mathcal{D}_{gl} \right] \delta \Pi^l_{\ mn}\;. \end{split} \end{align}
By defining \begin{equation} \label{khat equation}
\hat \cK_a^{\;\;b g r} =  \cK^{ \bar a}_{\ \bar b \bar g \bar r} \mathcal{G}_{a \bar a}^{\ \ \ b \bar b [g |\bar g| r] \bar r} \quad \text{and } g_\b=(g_b, \frac{1}{\l}) \;,  
\end{equation}
where the sums are restricted to $\mathcal M$ coordinates, 
the variation can be written succinctly as
\begin{align}
        \delta S = 0 =& \int d^\rd x \sqrt{|g|} \left[ E_a{}^{mn} - \tfrac{1}{\rd+1} \delta_a{}^{(m}E_{b}{}^{n)b}  \right] \delta \Pi^a{}_{mn} \\
        E_a{}^{mn} = E_a{}^{nm} =& \frac{1}{2\kappa_0 J_0 c}\breve\nabla_a\left( \sqrt{|g|}g^{mn} \right)-\breve\nabla_c\left( \sqrt{|g|} \hat \cK_{a}^{\,\,\, (m n) c} \right) \quad \nonumber\\
        &\quad +2 \l_0^2\breve\nabla_c\left( \sqrt{|g|}g_a \breve{\mathcal{K}}^{(mn)c}\right) \cr
        &\quad -2  \l_0^2\sqrt{|g|}K^{(mn)c}\cD_{ca} ~~~.
\end{align}
Here, $\breve \nabla_a$ is the derivative operator with respect to the fundamental projective invariant $\Pi^l_{\ mn}$ with action as follows:
\begin{align}
\breve{\nabla}_d\sqrt{|g|}=\partial_d \sqrt{|g|}-\Pi\indices{^a _{ad}}\sqrt{|g|}=\partial_d \sqrt{|g|}
\end{align}
\begin{align}
         \breve\nabla_d\left( \sqrt{|g|} \hat \cK_{a}^{\,\,\, m n c} \right) =& 
        (\partial_d \sqrt{|g|})   \hat \cK_{a}^{\,\,\, m n c}  + \sqrt{|g|}  \breve\nabla_d \hat \cK_{a}^{\,\,\, m n c} \cr
        =& (\partial_d \sqrt{|g|})   \hat \cK_{a}^{\,\,\, m n c}  + \sqrt{|g|} \left( \partial_d \hat \cK_{a}^{\,\,\, m n c} -\Pi^f{}_{ad} \hat  \cK_{f}^{\,\,\, m n c} \right) \cr
        & +\sqrt{|g|} \left( \Pi^m{}_{fd}\hat  \cK_{a}^{\,\,\, f n c}+  \Pi^n{}_{fd}\hat \cK_{a}^{\,\,\, m f c} +  \Pi^c{}_{fd}\hat \cK_{a}^{\,\,\, m n f}  \right)  .
\end{align}
Thus, the field equations for $\Pi^a{}_{mn}$ are
\begin{align}
        E_a{}^{mn} - \tfrac{1}{\rd+1} \delta_a{}^{(m}E_{b}{}^{n)b}  = 0~~~.
\label{Pi Field equations}\end{align}

 We note that if the connection were chosen to be compatible with the metric $g_{ab}$, then in the language of Tractor Calculus\cite{Eastwood2}, Eq.~(\ref{Pi Field equations})
would imply that the projective curvature is Yang-Mills \cite{Gover:2014vxa}. 
\subsection{Equations of motion for \texorpdfstring{\(\mathcal{D}_{bc}\)}{D} }

To find the field equations for \(\mathcal{D}_{bc}\), we proceed in the same manner as we did for \(\Pi^a_{\ bc}\). The contributions are of the form:
\begin{align} \label{eq:P Variation 1} \begin{split} S = \int \cK^a_{\ bcd}\mathcal{F}& \\ \implies \delta S &= \int \left(\delta \cK^a_{\ bcd}\right)\mathcal{F} \\ &= \int \delta\left(\mathcal{R}^a_{\ bcd} + \delta^a_{\ c}\mathcal{D}_{db} - \delta^a_{\ d}\mathcal{D}_{cb}\right)\mathcal{F} \\ &= \int \left[\left(\delta^a_{\ c}\delta^p_{\ d}\delta^q_{\ b}-\delta^a_{\ d}\delta^p_{\ c}\delta^q_{\
 b}\right)\mathcal{F}\right]\delta\mathcal{D}_{pq} \end{split} \end{align}
and
\begin{align} \label{eq:P Variation 2} \begin{split} S = \int \breve{\mathcal{K}}_{bcd}\mathcal{F}& \\ \implies \delta S &= \int  \left(\delta \breve{\mathcal{K}}_{bcd}\right)\mathcal{F} \\ &= \int \delta \left(\partial_{[c}\mathcal{D}_{d]b} + \Pi^e_{\ b[d}\mathcal{D}_{c]e}\right)\mathcal{F} \\ &= \int \left[ -\delta^p_{\ d}\delta^q_{\ b}\partial_c \mathcal{F} + \delta^p_{\ c}\delta^q_{\ b}\partial_d \mathcal{F} \right. \\ & \ \ \ \ \left. + \left( \Pi^q_{\ bd}\delta^p_{\ c} - \Pi^q_{\ bc}\delta^p_{\ d} \right) \mathcal{F} \right] \delta\mathcal{D}_{pq} \end{split} \end{align}
Again, by assigning the appropriate object \(\mathcal{F}\) to each term we have:
\begin{align} \label{eq:PEH P Variation} \begin{split} \delta S_{\text{PEH}} &= -\frac{1}{2\kappa_0} \int d^{\rm{d}}x \sqrt{|g|} \left[ (\rd-1) g^{qp} \right]\delta\mathcal{D}_{pq} \end{split} \end{align}
\begin{align} \label{eq:PGB1 P Variation} \begin{split} \delta S_{\text{PGB1}} &= -2 J_0 c \int d^{\rm{d}}x \sqrt{|g|} \cK^e_{\ fgh} \left[ \mathcal{B}_{ce}^{\ \ qfcgph} - \mathcal{B}_{ce}^{\ \ qfpgch} \right. \\ & \hspace{3.0cm} \left. + 2\lambda_0^{\ 2} g_c g_e g^{qf}g^{pg}g^{ch} \right]\delta\mathcal{D}_{pq} \end{split} \end{align}
\begin{align} \label{eq:PGB2 P Variation} \begin{split} \delta S_{\text{PGB2}} &= 4 J_0 c \lambda_0^{\ 2}  \int d^dx \left[ \partial_c \left( \sqrt{|g|} \breve{\mathcal{K}}_{fgh} g^{qf}g^{pg}g^{ch} \right)\right. \\ & \hspace{2.0cm} \left. + \sqrt{|g|} \breve{\mathcal{K}}_{fgh} \Pi^q_{\ bc} g^{bf} g^{pg} g^{ch} \right] \delta\mathcal{D}_{pq} \end{split} \end{align}
\begin{align} \label{eq:PGB3 P Variation} \begin{split} \delta S_{\text{PGB3}} &= 4 J_0 \lambda_0^{\ 2} \int d^{\rm{d}}x \left[ \sqrt{|g|} \breve{\mathcal{K}}_{fgh} g_a g^{qf}g^{ag}g^{ph} \right. \\ & \hspace{2.0cm} + \partial_{g}\left( \sqrt{|g|}\cK^{a}_{\ bcd}g_a g^{bq}g^{cp}g^{dg} \right) \\ & \hspace{2.0cm} \left. + \sqrt{|g|}\cK^a_{\ bcd}g_a g^{bf}g^{cp}g^{dg}\Pi^q_{\ fg} \right] \delta \mathcal{D}_{pq}\;. \end{split} \end{align}
Then the variation with respect to $\delta \mathcal{D}_{pq} $ yields  
\begin{align} \label{D Field equations}
-\frac{1}{2\kappa_0 J_0 c}& \sqrt{|g|} ({\rm{d}}-1) g^{pq} +\sqrt{|g|} \hat \cK_{c}^{\,\,\, (pq)c}  \nonumber\\
&+ 2 \l_0^2 \breve\nabla_g( \sqrt{|g|} K^{(pq)g} ) - 2 \l_0^2 \sqrt{|g|} g_c \breve{\mathcal{K}}^{(pq)c} =0\;.
\end{align}
Note the derivative of \(K\indices{^{(pq)g}}\), which make the field equations second-order differential equations in \(\mathcal{D}_{bc}\).

\subsection{Equations of motion for \texorpdfstring{\(g_{bc}\)}{g}}\label{Subsec:Metric}

Finally, we will find the field equations for the spacetime metric tensor \(g_{bc}\). For the sake of familiarity,  we will write the TW action $S$, Eq.(\ref{eq:PGB Action Rescaled}),  as 
\begin{equation}
S = -\frac{1}{2\kappa_0} \int d^{\rm{d}}x \sqrt{|g|} \ g^{ab} R(\G)_{ab} + \int d^{\rm{d}}x \sqrt{|g|}\, \mathcal{L}_\text{S}\  
\end{equation}
so as to separate the  Einstein-Hilbert-Palatini action from the rest of the action.  The  $\mathcal{L}_\text{S} $ is the remaining part of the Lagrangian density on $ \mathcal{M}$, viz., 
\begin{align} 
\mathcal{L}_{\text{S}}=&  - \frac{1}{2\kappa_0} ({\rm{d}}-1)\cP + c J_0 \lambda_0^{\ 2} K_{bcd}K^{bcd} 
-  c J_0  \left(\cK^{a}_{\ bcd}\cK_{a}^{\ bcd} - 4\cK_{ab}\cK^{ab} + \cK^2\right).
\end{align}
We have explicitly written the Ricci tensor as $R(\G)_{ab}$ to emphasize the independence of the connection from the metric.  In what follows  we write $R\equiv R(\G)_{ab} g^{ab}$ and $R_{ab} \equiv R(\G)_{ab}$.  The total action that contains the TW action and any matter fields is written as 
\begin{equation}
S_\text{total} = -\frac{1}{2\kappa_0} \int d^{\rm{d}}x \sqrt{|g|} \ g^{ab} R_{ab} + \int d^{\rm{d}}x \sqrt{|g|}\, \mathcal{L}_\text{S}+ \int d^{\rm{d}}x \sqrt{|g|}\, \mathcal{L}_\text{matter}.\  
\end{equation}
  We can define energy-momentum tensors $\Theta^\text{S}_{pq}$  and \( \Theta^\text{matter}_{pq}\) from the variation of the action with respect to the inverse metric \(g^{pq}\), via 
\begin{align} \label{Definition of Energy-Momentum Tensor}
\delta S_\text{total} =  \int \sqrt{|g|}& d^\rd x \bigg(\frac{1}{ \sqrt{|g|}}\frac{\delta(\sqrt{|g|}\mathcal L_\text{S})}{\delta g^{pq}} + \frac{1}{ \sqrt{|g|}}\frac{\delta(\sqrt{|g|}\mathcal L_\text{matter})}{\delta g^{pq}} \nonumber\\
& -\frac{1}{2\kappa_0 }\bigg(\frac{\delta R}{\delta g^{pq}} + \frac{R}{\sqrt{|g|}}\frac{\delta \sqrt{|g|}}{\delta g^{pq}}\bigg)\bigg)\delta g^{pq} =0\;.
\end{align}  Then from the  Einstein-Palatini equations, 
\begin{align}
\label{e:Einstein-Palatini_eqn}
 \frac{1}{2}R_{(pq)} -\frac{1}{2}R g_{pq} =\kappa_0 (\Theta^\text{S}_{pq}+ \Theta^\text{matter}_{pq}) \;,
\end{align} 
gives the  energy-momentum tensors  defined as,
 \be \Theta^\text{S}_{pq} =\frac{2}{ \sqrt{|g|}}\frac{\delta\big(\sqrt{|g|} \mathcal{L}_{\text{S}}\big)}{\delta g^{pq}}\,\, \;\text{and}\;\,\,\Theta^\text{matter}_{pq}=\frac{2}{ \sqrt{|g|}}\frac{\delta(\sqrt{|g|}\mathcal L_\text{matter})}{\delta g^{pq}}\;, \label{eq:Energy-Momentum} \ee
with $\kappa_0 =\frac{8 \pi G}{c^4}$.  Because the connection is not compatible with the metric, the left and right hand sides of Eq.(\ref{e:Einstein-Palatini_eqn}) are not separately divergence free.  We will address  this in the next subsection.  

More precisely,  the energy-momentum tensor, $\Theta^\text{S}_{pq}$, arises from $S$, by first exposing the  Einstein -Palatini tensor from the action which resides in the  \(S_{\text{PEH}}\) summand of $S$.  It  has the variation
\begin{align} \label{eq:PEH g Variation} \begin{split} \delta S_{\text{PEH}} &= -\frac{1}{2\kappa_0} \int d^{\rm{d}}x \cK^a_{\ bcd} \delta^c_{\ a} \delta\left(\sqrt{|g|}g^{bd}\right) \\ &= -\frac{1}{2\kappa_0} \int d^{\rm{d}}x \sqrt{|g|} \cK^a_{\ bcd} \delta^c_{\ a} \left( \delta^b_{\ i}\delta^d_{\ j} - \frac{1}{2} g_{ij}g^{bd} \right) (\delta g^{ij}) \\ &= -\frac{1}{2\kappa_0} \int d^{\rm{d}}x \sqrt{|g|} \left( \cK_{ij} - \frac{1}{2}g_{ij}\cK \right) (\delta g^{ij})\quad  \\ &= -\frac{1}{2\kappa_0} \int d^{\rm{d}}x \sqrt{|g|} \bigg( (R_{ij} - \frac{1}{2}g_{ij}R)\\
&\hspace{3.0cm} -({\rm{d}}-1)(\cP_{ij} -\frac{1}{2} \cP g_{ij}) \bigg) (\delta g^{ij})\;
 \end{split} \end{align}
where the Einstein-Palatini tensor is easily  recognized. Continuing to the \(S_{\text{PGB1}}\) term, we first find
\begin{align} \label{eq:g Variation 1} \begin{split} & \delta\left(\sqrt{|g|}\mathcal{B}_{ae}^{\ \ bfcgdh}\right) \mathcal{F} = \int \sqrt{|g|} \Big[-\frac{1}{2} g_{ij}\mathcal{B}_{ae}^{\ \ bfcgdh} \\ & \hspace{2.5cm} \left. + \delta^{c}_{\ a} \delta^{g}_{\ e} \delta^{b}_{\ i} \delta^{d}_{\ j} g^{f h} + \delta^{c}_{\ a} \delta^{g}_{\ e} \delta^{f}_{\ i} \delta^{d}_{\ j} g^{bd} \right. \\ & \hspace{2.5cm} \left. - 4 \delta^{c}_{\ a} \delta^{g}_{\ e} \delta^{b}_{\ i} \delta^{f}_{\ j} g^{dh} - 4 \delta^{c}_{\ a} \delta^{g}_{\ e} \delta^{d}_{\ i} \delta^{h}_{\ j} g^{bf} \right. \\ & \hspace{2.5cm} \left. - g_{ai} g_{ej} g^{bf} g^{cg} g^{dh} + g_{ae} \delta^{b}_{\ i} \delta^{f}_{\ j} g^{cg} g^{dh} \right. \\ & \hspace{2.5cm} + g_{ae} \delta^{c}_{\ i} \delta^{g}_{\ j} g^{bf} g^{dh} + g_{ae} \delta^{d}_{\ i} \delta^{h}_{\ j} g^{bf} g^{cg} \Big] \mathcal{F} (\delta g^{ij})\;. \end{split} \end{align}
We can get part of the variation of \(S_{\text{PGB1}}\) by putting a constant in front of Eq.~(\ref{eq:g Variation 1}) and plugging in the appropriate \(\mathcal{F} = \cK^{a}_{\ bcd} \cK^{e}_{\ fgh}\). The other part of the variation of \(S_{\text{PGB1}}\) can be found separately. Altogether, we have
\begin{align} \label{eq:PGB1 g Variation} \begin{split} & \delta S_{\text{PGB1}} = -J_{0} c \int d^{\rm{d}}x \sqrt{|g|} \Bigg[ -\frac{1}{2} g_{ij} \mathcal{B}_{ae}^{\ \ bfcgdh} \cK^{a}_{\ bcd} \cK^{e}_{\ fgh} \\ & \hspace{2cm} + 2 \cK_{ij}\cK - 8 \cK_{ib}\cK_{jd}g^{bd} - \cK_{iabd}\cK_{j}^{\ adb} \\ & \hspace{2cm} + \cK^c_{\ idb}\cK_{cj}^{\ \ db} + \cK^{ca}_{\ \ ib}\cK_{caj}^{\ \ \ b} + \cK^{cad}_{\ \ \ i}\cK_{cadj} \\ & \hspace{1.5cm} + \lambda_0^{\ 2} \Bigg\{ \frac{1}{{\rm{d}}+1} g_{ij} \partial_a \left(g_e g^{bf}g^{cg}g^{dh}\cK^{a}_{\ bcd}\cK^{e}_{\ fgh}\right) \\ & \hspace{1.2cm} - g_a g_e g^{cg}g^{dh}\Big( \frac{1}{2}g^{bf}g_{ij}\cK^{a}_{\ bcd}\cK^{e}_{\ fgh} + \cK^{a}_{\ icd}\cK^{e}_{\ jgh} \\ & \hspace{1.5cm} + \cK^{a}_{\ cdi}\cK^{e}_{\ ghj} + \cK^{a}_{\ dic}\cK^{e}_{\ hjg}  \Big) \Bigg\} \Bigg] (\delta g^{ij})\;. \end{split} \end{align}
Similarly, variation of \(S_{\text{PGB2}}\) is given by
\begin{align} \label{eq:PGB2 g Variation} \begin{split} \delta S_{\text{PGB2}} &= J_0 \lambda_0^{\ 2} \int d^{\rm{d}}x \ \delta\left(\sqrt{|g|}g^{bf}g^{cg}g^{dh}\right)\breve{\mathcal{K}}_{bcd}\breve{\mathcal{K}}_{fgh} \\ &= J_0 \lambda_0^{\ 2} \int d^{\rm{d}}x \sqrt{|g|} \left[ \frac{1}{2} g_{ij} g^{bf}g^{cg}g^{dh} + \delta^{c}_{\ i}\delta^{g}_{\ j} g^{dh}g^{bf} \right. \\ & \left. + \delta^{d}_{\ i}\delta^{h}_{\ j} g^{cg}g^{bf} + \delta^{b}_{\ i}\delta^{f}_{\ j} g^{cg}g^{dh} \right] \breve{\mathcal{K}}_{bcd} \breve{\mathcal{K}}_{fgh} (\delta g^{ij}) \\ &= J_0 \lambda_0^{\ 2}  \int d^{\rm{d}}x \sqrt{|g|}  g^{cg}g^{dh} \Big( \frac{1}{2}g^{bf}g_{ij}\breve{\mathcal{K}}_{bcd}\breve{\mathcal{K}}_{fgh} \\ &  + \breve{\mathcal{K}}_{cdi}\breve{\mathcal{K}}_{ghj} + \breve{\mathcal{K}}_{dic}\breve{\mathcal{K}}_{hjg} + \breve{\mathcal{K}}_{icd}\breve{\mathcal{K}}_{jgh} \Big)(\delta g^{ij})\;.  \end{split} \end{align}
Finally, the variation of \(S_{\text{PGB3}}\) is given by
\begin{widetext}
\begin{align} \label{eq:PGB3 g Variation} \begin{split} \delta S_{\text{PGB3}} &=  2J_0 \lambda_0^{\ 2}  \int d^{\rm{d}}x \ \delta\left(\sqrt{|g|}g_a g^{bf}g^{cg}g^{dh}\right)\cK^{a}_{\ bcd}\breve{\mathcal{K}}_{fgh} \\ &=  2 J_0 \lambda_0^{\ 2} \int d^{\rm{d}}x \sqrt{|g|} \Bigg[ g_a g^{cg}g^{dh}\Big( \cK^a_{\ icd}\breve{\mathcal{K}}_{jgh} \\ & \hspace{5cm} + \cK^a_{\ cdi}\breve{\mathcal{K}}_{ghj} + \cK^a_{\ dic}\breve{\mathcal{K}}_{hjg} \Big) \\ & \hspace{3cm} - \frac{1}{2({\rm{d}}+1)} g_{ij} \partial_a\left( g^{bf}g^{cg}g^{dh}\cK^{a}_{\ bcd}\breve{\mathcal{K}}_{fgh} \right) \Bigg] \left(\delta g^{ij}\right). \end{split} \end{align}
\end{widetext}
Putting this all together defines the TW energy-momentum tensor $\Theta^{\text{S}}_{ij}$ as 
\begin{widetext}
\begin{align} \label{TW Energ-Momentum Tensor} \begin{split}  \Theta^{\text{S}}_{mn} =& -\tfrac{\rd-1}{2\kappa_0} \cP_{(mn)} + g_{mn} \left[2 J_0 c\l_0^2 (-\tfrac{1}{\rd+1} \nabla_a + \Delta_a) \cK^{a}{}_{bcd}K^{bcd}-\mathcal{L}_{\text{S}}\right] \cr
       &+ 2J_0c \lambda_0^2 (K_{mcd}K_n{}^{cd} + 2 K_{bcm}K^{bc}{}_n) + 2 J_0 c(8 \cK_{mb}\cK^b{}_n - 2 \cK \cK_{mn})\cr
       &+ 2 J_0 c (\cK_{mbcd}\cK_n{}^{bcd} -\cK^a{}_{mcd}\cK_{an}{}^{cd} - 2 \cK^{abc}{}_m \cK_{abcn})
\end{split} \end{align}
\end{widetext}
where \(K_{bcd} = g_a \cK^a_{\ bcd} + \breve{\mathcal{K}}_{bcd}\) and $\Delta_a\equiv g_a-\alpha_a$ are tensors. This demonstrates that the energy-momentum tensor is indeed manifestly tensorial on \(\mathcal M\).
Since we have used the Gauss-Bonnet action to describe dynamics for $\cD_{ab}$, the field  equations are second-order differential equations in \(g_{ab}\).

\subsection{Palatini Field and Metric Compatible Connection }\label{Subsec:Palatini}
It may be convenient to solve the field equations with a connection that is the Levi-Civita connection of a metric.  Here we demonstrate how we may exchange the field degrees of $\Pi^a_{\;\;bc}$ for a tensor, $C^a_{\;\;bc}$  which we will call the \emph{Palatini field}, and a Levi-Civita connection associated with the metric $g_{ab}$. 
Given any two connections, say $\Gamma^a{}_{bc}$ and $\hat \Gamma^a{}_{bc}$, their difference is always a tensor. Define $\hat \Gamma^a{}_{bc}$ as the Levi-Civita connection associated with $g_{ab}$ so that $\hat \nabla_a g_{bc} =0$. Then, a \emph{Palatini field}  $C^a_{\;\;bc}$, can be defined relative to this Levi-Civita connection for any connection $\Gamma^a{}_{bc} $ as 
\begin{equation}
\Gamma^a{}_{bc} = \hat \Gamma^a{}_{bc} + C^a_{\;\;bc}. \label{Palatini Field}
\end{equation}
Here $ C^a_{\;\;bc}=  C^a_{\;\;cb}$ as there is no torsion.  Similarly for the projective invariant, $\Pi^a{}_{bc} $, we may write 
\begin{equation}
\Pi^a{}_{bc} = \hat \Pi^a{}_{bc} + \tilde C^a_{\;\;bc}, \label{Projecitve_Palatini_Field}
\end{equation} 
where $ \hat \Pi^a{}_{bc}$ is the projective invariant for the equivalence class in which $\hat \Gamma^a{}_{bc}$ is a member and $\tilde C^a_{\;\;bc} \equiv  C^a_{\;\;bc}-\frac{1}{\rd+1} (\delta^{a}_{\;\;c}~ C_{b}+\delta^{a}_{\;\;b}~ C_{c}), $ is trace-free and symmetric in its last two indices.  Here $C_{b} \equiv  C^a_{\;\;ba} $.     

Using the Palatini field and the Levi-Civita connection, the Riemann curvature tensor for $\Gamma^a{}_{bc}$ may be written as 
\begin{eqnarray}
R^m_{\,\,\,nab} &=& \hat R^m_{\,\,\,nab} + \hat \nabla_a C^m_{\;\;\;nb}- \hat \nabla_b C^m_{\;\;\,na}+  C^r_{\;\;nb} C^m_{\;\;\,ar}- C^r_{\;\;na} C^m_{\;\;\;br} \\&\equiv& \hat R^m_{\,\,\,nab} + Q^m_{\,\,\,nab}. 
\end{eqnarray}
Similarly, 
\be
R_{ab} = \hat R_{ab} + Q_{ab},
\ee
where $Q_{ab} \equiv Q^m_{\,\,\,amb}$ and $Q= g^{ab} Q_{ab}$.
Then the LHS\ of Eq. (\ref{e:Einstein-Palatini_eqn}) may be written as 
\begin{equation}
\frac{1}{2}R_{(ab)} - \frac{1}{2} R g_{ab} = \hat R_{ab} - \frac{1}{2} \hat R g_{ab} + \frac{1}{2}Q_{(ab)} - \frac{1}{2} Q g_{ab}.  
\end{equation}
In leu of the field variables $\{ g_{ab}, \Pi^r_{\;\;na}, \mathcal{D}_{ab}\}$, the field equations  may now be solved using the fields $\{g_{ab}, C^r_{\;\;na}, \mathcal{D}_{ab}\}  $ and Eq. (\ref{e:Einstein-Palatini_eqn}) becomes 
\be  \hat R_{pq} -\frac{1}{2}\hat R g_{pq} =\kappa_0 (\Theta^\text{S}_{pq}+ \Theta^\text{matter}_{pq}) + \frac{1}{2}Q g_{pq} - \frac{1}{2}Q_{(pq)} \label{Divergence}. \ee In this way, both sides of Eq. (\ref{Divergence}) are separately divergence-free with respect to the Levi-Civita connection. One sees that both $C^r_{\;\;nb}$ and $\mathcal{D}_{ab}$ act as  geometric sources for the metric compatible Riemannian geometry in general relativity.  When $\mathcal{P}_{ab}=0$  and $C^r_{\;\;nb}=0$ this becomes the usual theory of general relativity.  Note in this case,  Eq. (\ref{P_in_terms_of_D}) becomes
\begin{align}
\mathcal{D} _{bc} =& \partial_b g_c - \hat \Gamma^e{}_{bc}g_e - g_b g_c,
\end{align}
which can be eliminated by a choice of coordinates (volume preserving). The analogy of $\mathcal{D}_{ab}$ with vector potentials $A_a$ in Yang-Mills theories  \cite{Lano:1994gx,Rodgers:1994ck,Rai:1989js,Rodgers:2003an} demonstrates that general relativity is in the ``pure gauge" sector of TW gravity. This strategy facilitates finding out whether there are projective geometric contributions to, for example, primordial perfect fluids, the origin of an  inflaton and  dark matter sources that may not have arisen from the matter Lagrangian. For solutions associated with definite symmetries, one can choose an ansatz for $C^a_{\;\;bc}$ and $\mathcal{D}_{ab}$ whose Lie derivative with respect to the Killing vectors of the metric vanish.   Recent work \cite{GoverA.Rod2012DEgE,Brensinger:2019mnx} has already shown that  projective geometry   serves as a  source for the cosmological constant.  Other issues related to the principle of equivalence, cosmology,  holonomy and projectively equivalent manifolds have been studied as well \cite{Hall:2007wp,Hall:2009zza,Hall:2011zza}.     
%
\section{Geodesic Deviation}

To complete this study of the gauge covariant field equations and gauge invariant action we examine the \textit{geodesic deviation} equations on the Thomas Cone and their image on the manifold $\mathcal M$. Not only does geodesic deviation have importance in tidal forces, it can also provide a mechanism to study radiative degrees of freedom in $\cD_{ab}$. Here, we will examine the modification to geodesic deviation that results from the presence of the projective gauge field \(\mathcal{D}_{bc}\).
A review of geodesic deviation and its derivation in general relativity can  be found in textbooks such as\cite{Hobson:2006se}.

\subsection{The geodesic deviation equation}
Let \(\mathcal M\) be the spacetime manifold equipped  with a metric  \(g_{ab}\). Recall the geodesic equation for any connection \(\Gamma^a_{\ bc}\)  on $\mathcal M$ 
\begin{equation} \label{eq:Geodesic Equation} \frac{d^2x^a}{d\tau^2} + \Gamma^{a}_{\ bc}\frac{dx^b}{d\tau}\frac{dx^c}{d\tau} = f(\tau)\frac{dx^a}{d\tau}\;, \end{equation}
where \(\tau\) is some parameter.  Here \(f(\tau)=0  \)   if and only if \(\tau\) is an affine parameter for \(\Gamma^a_{\ bc}\). In the presence of a gravitational field where the connection $\Gamma^a_{\ bc}$ is compatible with the metric, freely moving objects will travel along geodesics specified by Eq.~(\ref{eq:Geodesic Equation}).

Consider the space of geodesics \(x^a(s,\tau)\), where for each fixed value \(s=s_0\), we have that \(x^a(s_0,\tau)\) is a geodesic with affine parameter \(\tau\). This gives us a one-parameter family of geodesics which allows us to examine geodesics that are close to each other. The geodesic tangent vector \(T^a(s,\tau)\) and geodesic deviation vector \(X^a(s,\tau)\) are given by
\begin{align} \label{eq:Tangent and Deviation Vectors} \begin{split} T^a(s,\tau) &= \frac{\partial x^a(s,\tau)}{\partial \tau}, \\ X^a(s,\tau) &= \frac{\partial x^a(s,\tau)}{\partial s}\;. \end{split} \end{align}
Eq.~(\ref{eq:Tangent and Deviation Vectors}) leads to an immediate relation between derivatives of \(T^a\) and \(X^a\)
\begin{equation} \label{eq:T and X Derivatives} \frac{\partial X^a}{\partial \tau} = \frac{\partial T^a}{\partial s}\;. \end{equation}
For a vector field \(V^a\) on \(\mathcal M\), the \textit{intrinsic derivative} of \(V^a\) along a curve \(x^a(\tau)\) is given by 
\begin{equation} \label{eq:Intrinsic Derivative} \frac{D V^a}{d \tau} = \nabla_{T^b \frac{\partial}{\partial x^b}}V^{a} = T^b\;. \nabla_b V^a \end{equation}
Using Eq.~(\ref{eq:Intrinsic Derivative}), we can find an acceleration by taking the second intrinsic derivative of a vector field. If we do this with the geodesic deviation vector \(X^a(s,\tau)\) with respect to \(\tau\), we find
\begin{align} \label{eq:Intrinsic Acceleration} \begin{split} \frac{D^2 X^a}{\partial \tau^2} &= T^c\nabla_c(T^b\nabla_b X^a) \\ &= \frac{\partial^2 T^a}{\partial s \ \partial \tau} + (\partial_c \Gamma^a_{\ bd})T^cT^bX^d \\ & \ \ \ \ + \Gamma^a_{\ bd} \left(\frac{\partial T^b}{\partial \tau} X^d + T^b \frac{\partial X^d}{\partial \tau} \right) + \Gamma^a_{\ cd} \bigg( \frac{\partial T^d}{\partial \tau} \\
& \ \ \ \ + \Gamma^d_{\ be} T^b X^e \bigg)T^c\;. \end{split} \end{align}
Eq.~(\ref{eq:Intrinsic Acceleration}) can be simplified since \(x^a(s,\tau)\) is a geodesic curve for all fixed \(s\). Due to this fact, we know that
\begin{align} \label{eq:Geodesic Intrinsic Derivative} \begin{split} &T^b\nabla_bT^a = 0 \\ \implies &X^c\nabla_c(T^b\nabla_bT^a) = 0\;. \end{split} \end{align}
Expanding Eq.~(\ref{eq:Geodesic Intrinsic Derivative}) and rearranging terms yields
\begin{align} \label{eq:Intrinsic Derivative Intermediate Step} \begin{split} \frac{\partial^2 T^a}{\partial s \ \partial \tau} = &-(\partial_d\Gamma^a_{\ cb})T^cT^bX^d \\ &-\Gamma^a_{\ bd} \left( \frac{\partial T^b}{\partial s}T^d + \frac{\partial T^d}{\partial s}T^b \right) \\ &-\Gamma^a_{\ cd}\left( \frac{\partial T^d}{\partial \tau} + \Gamma^d_{\ be} T^b T^e \right)X^c\;. \end{split} \end{align}
Using Eq.~(\ref{eq:Intrinsic Derivative Intermediate Step}), we eliminate \(\frac{\partial^2 T^a}{\partial s \ \partial \tau} \) \ from Eq.~(\ref{eq:Intrinsic Acceleration}) and find
\begin{align} \label{eq:Geodesic Deviation} \begin{split} \frac{D^2X^a}{\partial \tau^2} &= \left(\partial_c\Gamma^a_{\ db} - \partial_d\Gamma^a_{\ cb} + \Gamma^a_{\ ce}\Gamma^{e}_{\ db} - \Gamma^a_{\ be}\Gamma^e_{\ cb} \right)T^cT^bX^d \\ &= R^a_{\ bcd}T^bT^cX^d\;. \end{split} \end{align}
This is the geodesic deviation equation.
Note we did not use metric compatibility to arrive at this expression. 
The full  Riemann curvature tensor \(R^{a}_{\ bcd}\)  appears in the geodesic deviation equation, including the Weyl term which does not usually appear in Einstein field equations. Gravitational radiation can influence  geodesic deviation directly making it a useful observational tool.  We will now explore the projective modifications of the geodesic deviation equation and insights on how the diffeomorphism field may be observed.  
\subsection{Projective geodesic deviation }

We turn our attention to the diffeomorphism field which we also may consider as the projective gauge field \(\mathcal{D}_{bc}\). To compute the resulting geodesic deviation on the spacetime manifold \(\mathcal M\) for a general connection, we  first must find the geodesic deviation of the TW connection on \(\mathcal N\), and project this deviation down onto \(\mathcal M\).

From Eq.~(\ref{eq:Geodesic Deviation}), the geodesic deviation \(X^{\alpha}(\tau)\) of the TW connection on \(\mathcal N\) is given by
\begin{equation} \label{eq:TW Geodesic Deviation} \frac{D^2X^{\alpha}}{d\tau^2} = \cK^{\alpha}_{\ \beta \sigma \rho} \frac{dx^{\beta}}{d\tau}\frac{dx^{\sigma}}{d\tau}X^{\rho}\;, \end{equation}
where the Greek indices range over all coordinates on \(\mathcal N\).  Now, as in Eq.~(\ref{General Projected Vector}) let \[
X^\a = (X^a, -\l X^a g_a+X^{5})\] define the projective geodesic deviation vector.  We have included a perpendicular component as physical vectors such as $X^\a = \Psi \tilde \g^\a\Psi$ might arise.  However, for simplicity we will ignore the $X^5$ component in this discussion. We have used  $g_a$  defined via a metric on $\mathcal N$ so as not to spoil the projective covariance of the equation. Let us first consider the  geodesic deviation \(X^a\) where \(a\) is a spacetime manifold coordinate specifically (not \(\lambda\)). Since  the only non-vanishing components of \(\cK^{\alpha}_{\ \beta \sigma \rho}\) are the components \(\cK^{\lambda}_{\ b c d}\) and \(\cK^{a}_{\ bcd}\), then Eq.~(\ref{eq:TW Geodesic Deviation}) reduces for \(\alpha = a\) to
\begin{align} \label{eq:TW Geodesic Deviation on Manifold} \begin{split} \frac{D^2X^a}{d\tau^2} &= K^a_{\ b c d} \frac{dx^b}{d\tau}\frac{dx^c}{d\tau}X^d \\ &= \left( \mathcal{R}^a_{\ bcd} + \delta^{a}_{\,[c}\mathcal{D}_{d]b} \right) \frac{dx^b}{d\tau}\frac{dx^c}{d\tau}X^d \\ &= \left( R^a_{\ bcd} + \delta^{a}_{\,[c}\mathcal{P}_{d]b} - \delta^a_{\ b}\mathcal{P}_{[cd]} \right) \frac{dx^b}{d\tau}\frac{dx^c}{d\tau}X^d \;. \end{split} \end{align}
Here $R^a_{\,\,\,\,bcd}$ is the Riemann curvature tensor for a connection $\G^a_{\,\,\,b c}$ which is not necessarily   compatible with the metric defining $g_a$. Now the parameter  \(\tau\) is an affine parameter \textit{for the TW connection on $\mathcal N$}, not for the $\G^a_{\,\,\,b c}$ connection on \(\mathcal M\). If we make a change of parameterization \(\tau \to u\) so that \(u\) is an affine parameter for the spacetime manifold connection,  we get using Eq.~(\ref{u goes to tau})
\begin{align} \label{eq:TW Geodesic Deviation Affine Parameter}   &\frac{D^2X^a}{du} - \bigg( \mathcal{R}^a_{\ bcd} + \delta^{a}_{\,[c}\mathcal{D}_{d]b} \bigg) \frac{dx^b}{du}\frac{dx^c}{du}X^d= \bigg( \frac{2}{\lambda} \frac{d\lambda}{du} \bigg) \frac{DX^a}{du} \nonumber \\
& \implies \frac{D^2X^a}{du} - \bigg( R^a_{\ bcd} + \delta^{a}_{\,[c}\mathcal{P}_{d]b} - \delta^a_{\ b}\mathcal{P}_{[cd]} \bigg) \frac{dx^b}{du}\frac{dx^c}{du}X^d \nonumber\\
& \hspace{5cm} = \Big( \frac{2}{\lambda} \frac{d\lambda}{du} \Big) \frac{DX^a}{du} \;. 
\end{align}
If we consider the $\l$ component, we find
\begin{align} \label{eq:TW Geodesic Deviation for Lambda Component} \begin{split} &\frac{D^2X^\l}{d\tau^2} = K^\l_{\ b c d} \frac{dx^b}{d\tau}\frac{dx^c}{d\tau}X^d \\ &\frac{D^2 (-\l X^a g_a)}{d\tau^2}= \left( \partial_{[b} \mathcal{D}_{c]a}+ \Pi^d_{\,\,\, a [c}\mathcal{D}_{b]d} \right) \frac{dx^b}{d\tau}\frac{dx^c}{d\tau}X^d \\ &= \left(\nabla_{[c}\cP_{d] b} + \a_{[d} \cP_{c] b}+ \a_{[b} \cP_{c] d} -R^a_{\ bcd} \a^a \right) \frac{dx^b}{d\tau}\frac{dx^c}{d\tau}X^d \;. \end{split} \end{align}
If we take $g_a =0$ as a gauge choice, the left hand side of the above expression vanishes, leaving \begin{equation}
 \left(\nabla_{[c}\cP_{d] b} + \a_{[d} \cP_{c] b}+ \a_{[b} \cP_{c] d} -R^a_{\ bcd} \a^a \right) \frac{dx^b}{d\tau}\frac{dx^c}{d\tau}X^d=0\;.
\end{equation} This illustrates the complexity of  \(\mathcal{D}_{bc}\) (or equivalently, \(\mathcal{P}_{bc}\)) as a dynamical  field, since it has its own field equations and energy-momentum tensor.  \(\mathcal{D}_{bc}\) will interact with the spacetime geometry and have an effect on \(\mathcal{R}^a_{\ bcd}\). Thus  geodesic deviation is a valuable resource for observation, and the projective gauge field could explain defects in these observations via Eq.~(\ref{eq:TW Geodesic Deviation Affine Parameter}).

\section{Conclusion}

String theory may be thought of as originating from regulating Feynman diagrams in gravitational theories, by adding a tiny dimension to the point particle as initial data.  This regulator quickly takes on a life of its own through the Virasoro algebra, which maintains the reparameterization invariance. It has been shown \cite{Brensinger:2017gtb} that a projective structure and subsequent projective geometry are the ubiquitous concepts that give meaning to this reparameterization in any dimension.  In projective geometry, a manifold is geometrically classified in terms of its family of geodesics. In many ways, geodesics are  the most experimentally available geometric structures  that give physicists access to the underlying geometry of a manifold. Affine geodesic lines, whether space-like, time-like, or even null, enjoy reparameterization invariance irrespective of the underlying metric. Furthermore, the correspondence between the Virasoro algebra and projective geometry  is analogous to the correspondence of an affine Lie algebra (a class of Kac-Moody algebras) for one-dimensional gauge transformations to Yang-Mills vector potentials in higher dimensional field theories; see Table \ref{table:1}.

\begin{table}[t]
\centering
\begin{tabular}{ ||m{4cm}|| m{2.3cm}|  m{2.2cm} || m{2.3cm}| m{3cm} ||  }
 \hline
  \multicolumn{1}{||c||}{Symmetry} &\multicolumn{2}{|c|}{String Theory}& \multicolumn{2}{|c|}{Field Theory} \\
 \hline \hline
 Reparameterization Invariance &Algebra: Virasoro &Coadjoint Elements: $(B,q)$ &Connection: Projective&\  $\tilde\nabla_\a(\mathcal{D}_{bc}, \Pi^a_{\,\,bc})$\\
 \hline
Gauge Invariance & Algebra: Affine Lie & Coadjoint Elements: ($A$, $\a$)& Connection: Yang-Mills&\ $D_a({A}_b)$\\ 
\hline
\end{tabular}
\caption{\emph{Correspondence of Symmetries in String Theories to Connections in Field Theories}\\ Coadjoint elements of the Virasoro algebra, $(B,q)$, consists of a quadratic differential $B$ and a central element $q$. They are in correspondence with the projective connection components $\mathcal{D}_{ab}$ that appear in the  projective covariant derivative $\tilde\nabla_\a$.  Analogously,  the coadjoint elements of the affine Lie algebra (Kac-Moody algebra), $(A,\a)$,  consisting of a one form $A$ and a central element $\a,$  and are in correspondence with the Yang-Mills connection, ${A}_a$ that appears in the gauge covariant derivative $D_a$. }
\label{table:1}
\end{table}
The projective geometry of Thomas and later Whitehead \cite{Thomas:1925a,Thomas:1925b,Whitehead} allows us to form a gauge theory for unparameterized paths which induces a dynamical field called the diffeomorphism field.  These  projective connections get their dynamics from the Thomas-Whitehead Gravitational Action defined in \cite{Brensinger:2017gtb}. However,  those and subsequent results \cite{Brensinger:2019mnx} used specific coordinates such as constant volume coordinates and background metrics.  In this note, we present the full gauge invariant Thomas-Whitehead action.  There are many advantages of having a gauge invariant theory, including the understanding of spontaneously broken symmetry and the constraints that arise in classical and quantum field theories.  The results here show precisely how any Dirac fermion will interact with the diffeomorphism field and how chiral masses become manifest due to a volume scale. These gravitationally induced chiral masses are affected by the dimension of the manifold, the number of gauge fields and the spinor's tensor density.  

The use of geodesics extends far beyond gravitational theories and these results may be of value in fluid dynamics, optimization, other gauge theories and even quantum computing.  Several projects applying the general TW theory presented in this paper are currently underway including the quantization of the fully covariant TW theory, sourcing of cosmological inflation, constraints imposed by affects on gravitational radiation, and applications to the understanding of dark matter.

\section*{Acknowledgments}
The research of K.\ S.\ is supported in part by the endowment of the Ford Foundation Professorship of Physics at Brown University. K. S. would like to thank Kevin Iga and Konstantinos Koutrolikos  for helpful discussions.
V. R. thanks S. L. Gallon for discussions.
 The research of K. H. and S. B.  is supported by fellowships from the Graduate College  at The University of Iowa.

\appendix
\section{Units, Conventions, and Helpful Calculations}\label{a:Conventions}
The units of the various constants used throughout this paper for $\rd = 4$ are
\begin{align}
\begin{split}
        [J_0] = \frac{M L^2}{T}\;, \quad[\cD_{ab}] &= [R_{ab}] = L^{-2}\;, \cr
           \quad[\ell] =  \text{dimensionless}\;,& \quad \left[\kappa_0 \right] =\frac{T^2}{ML} \cr   [d^{\rd}x] =& T L^{\rd -1}
        \end{split}
\end{align}
We may at times set $c=1$ but expose factors of $c$ when calculating numerical values.
Latin indices take values $a,b,\dots = 0,1,2,\dots, \rd-1$ and Greek indices take values \\ $\mu,\nu,\dots = 0,1,2,\dots, \rd$, with the exception of the Greek letter $\lambda$, which refers to the projective coordinate $x^{\rd} = \lambda = \lambda_0 \ell$. A coordinate transformation and corresponding Jacobian matrix over the $\rd$-dimensional space is given as
\begin{align}
        x'^m = x'^m(x^n)~~~,~~~J^m{}_n  = \frac{\partial x'^m}{\partial x^n}
\end{align}
A useful property of the determinant of the Jacobian matrix is its derivative in terms of the coordinates:
\begin{align}
        \frac{\partial \log |J|}{\partial x^a} = \frac{\partial x^n}{\partial x'^m} \frac{\partial }{\partial x^a} \frac{\partial x'^m}{\partial x^n}.
\end{align}

Our conventions for the Riemann curvature tensor $R^{a}{}_{bcd}$ are the same as for the projective curvature $\cK^\mu{}_{\nu\alpha\b}$. The Riemann curvature tensor is written in terms of $\G^{m}{}_{ab}$ where as the projective curvature is written in terms of $\tilde{\G}^{\mu}{}_{\a\b}$:
\begin{align}\label{e:PCurvApp}
       \cK^\mu{}_{\nu\alpha\b} \equiv \tilde{\G}^\mu{}_{\nu[\b,\a]} + \tilde{\G}^\rho{}_{\n[\b}\tilde{\G}^\mu{}_{\a]\rho}~~~.
\end{align}
Here and throughout, brackets mean anti-symmetrization and parenthesis mean symmetrization
\begin{align}
\cK^{\alpha}{}_{\beta[\mu\nu]} =& \cK^{\alpha}{}_{\beta\mu\nu} - \cK^{\alpha}{}_{\beta\nu\mu} ~~~,~~~ \cK_{(\mu\nu)} = \cK_{\mu\nu} + \cK_{\nu\mu}\;.
\end{align}
Eq.~(\ref{e:PCurvApp}) means the following must be true

\begin{align} 
[{\tilde \nabla}_\a,{\tilde \nabla}_\b] V^\g &= \,{\cK}^{\g}_{\,\,\,\rho\a \b  } V^\rho \\
[{\tilde \nabla}_\a,{\tilde \nabla}_\b] V_\g &=- \,{\cK}^{\rho}_{\,\,\,\g\a \b  } V_\rho \;  .
\end{align}

The $\rd$-dimensional metric $g_{ab}$ is promoted to the Thomas cone metric $G_{\alpha\beta}$ by adding the appropriate projective contributions to the components.  An easy way to see this is by writing:
\begin{align}
       & G_{\a\b}  = \begin{bmatrix}
  g_{ab}-\lambda_0^{\ 2}g_ag_b & -\frac{\lambda_0^{\ 2}}{\lambda} g_a \\
  -\frac{\lambda_0^{\ 2}}{\lambda} g_b & -\frac{\lambda_0^{\ 2}}{\lambda^2}
  \end{bmatrix}\;. \label{e:GmetricGenapp} \\
   &   G^{\a \b} = \begin{bmatrix}
  g^{ab} & -\lambda g^{am}g_m \\
  -\lambda g^{bm}g_m & \frac{\lambda^2}{\lambda_0^{\ 2}}\left(-1 + g^{mn}\lambda_0^{\ 2}g_mg_n\right)
  \end{bmatrix} \;,\\
       &        G_{\a\b} = \delta^a_{\,\,\alpha}{} \delta^b_{\,\,\beta}{} \,g_{ab} - \lambda_0^2\, g_\alpha g_\b\\
       &G^{\a\b} = g^{ab} (\delta^\alpha_{\,\,\,\,a} - g_a \Upsilon^\a)(\delta^\b_{\,\,\,b} - g_b \Upsilon^\b) - \lambda_0^{-2} \Upsilon^\a\Upsilon^\b \label{e:deltaGmetricAndInverseapp} 
\end{align}
where the $\rd$-dimensional metric $g_{ab}$ has signature $(+,-,-,-,\cdots,-)$ and the dimensionless parameter $\ell = \lambda/\lambda_0$. The function $g_a \equiv \frac{1}{d+1} \partial_a \log{\sqrt{|g|} }$ is chosen as it transforms like the trace of a connection  and depends only on the metrics determinant.  The $\rd$-dimensional Riemann Curvature tensor $?R^a_bcd?$ satisfies the same relation as the \text{$(\rd+1)$}-dimensional tensor $?\cK^\alpha_\beta\mu\nu?$, Eq.~(\ref{e:Kaction}), but in terms of the $\rd$-dimensional covariant derivative $\nabla_a$. The commutator of covariant derivatives on an arbitrary rank $m$-covariant, rank $n$-contravariant tensor is equivalent to the following action of $?R^a_bcd?$
\begin{align}\label{e:RactionGeneral}
        &[\nabla_a , \nabla_b ] ?T_{c_1\dots c_m}^{d_1\dots d_n}? =\nonumber\\
         &\hspace{1cm}-?R^e_{c_1 ab}? ?T_{ec_2\dots c_m}^{d_1 d_2 \dots d_n}? - \dots -?R^e_{c_m ab}? ?T_{c_1 c_2\dots e}^{d_1 d_2 \dots d_n}? \nonumber\\
        &\hspace{1cm}+ ?R^{d_1}_{eab}? ?T_{c_1\dots c_m}^{e\dots d_n}? + \dots + ?R^{d_m}_{eab}? ?T_{c_1\dots c_m}^{d_1\dots e}? 
\end{align}
We list all non-vanishing connections and curvatures below:
\begin{align}
        \label{e:Gammaa}
        &\tilde{\G}^a{}_{bc} = \Pi^a{}_{bc}; \quad\tilde{\G}^\l{}_{ab} = \l \cD_{ab}\;,\\&\tilde{\G}^a{}_{\l b}=\tilde{\G}^a{}_{b \l} = \lambda^{-1} \delta^a_{\,\,\,b}{}\;,\\ 
                \label{e:PPi}
                 &{\Pi}^{a}_{\,\,\,\,b c} = { \G}^{a}_{\,\,\,\,b c} + \delta^{a}_{(\,c}~ \a_{b)}\\&                \mathcal{P}_{bc} = \mathcal{D}_{bc} - \partial_b \a_c + \Gamma^e_{\ bc}\a_e + \a_b \a_c\;\end{align} 

\begin{align}
        \label{e:KRiemanna}
        &\cK^a{}_{bcd} = R^a{}_{bcd} + \delta^{a}_{[\,\,c}{} \cP_{d]b}- \delta^a{}_b \mathcal{P}_{[cd]}\\
        &
 \cK^{a}_{\ bcd} = \mathcal{R}^{a}_{\ bcd} + \delta^a_{\ [c}\mathcal{D}_{d]b} \\ &\cK^{\lambda}_{\ cab} = \lambda \partial_{[a}\mathcal{D}_{b]c} + \lambda \Pi^{d}_{\ c[b}\mathcal{D}_{a]d}  \\&\cK^{\lambda}_{\ bcd} = \lambda ( \partial_{[c}\mathcal{P}_{d]b} + \Gamma^a_{\ b[d} \mathcal{P}_{c]a} + \alpha_{[d} \mathcal{P}_{c]b} \nonumber\\
& \hspace{2.0cm}+ \alpha_b\mathcal{P}_{[cd]} - R^a_{\ bcd}\alpha_a ) \\
& \breve{\mathcal{K}}_{bcd}\equiv \frac{1}{\lambda}\mathcal{K}\indices{^\lambda _{bcd}}\\
        \label{e:KRiccia}
        &\cK^{\mu}{}_{b\mu d} =\cK_{bd} = \mathcal{R}_{bd} + ({\rm{d}}-1)\mathcal{D}_{bd} \cr &\quad \quad\quad= R_{bd} + \rd\mathcal{P}_{db} - \mathcal{P}_{bd}  \\
        \label{e:Ka}
        &\cK\equiv  G^{\alpha\beta}\cK_{\alpha\beta}=\cR + (\rd - 1)\cD = R + (\rd - 1)\cP \\
        \label{e:RDa}
                &R = g^{ab}R_{ab}\;,\quad \cP =  g^{ab}\cP_{ab}\;.
\end{align}
\begin{align}  \begin{split}  K_{bcd} &\equiv  g_a\cK^a_{\ bcd} + \breve{\mathcal{K}}_{bcd}\\
 &=  \left( g_a-\alpha_a \right) R^a_{\ bcd} + \left(g_c-\alpha_c\right)\mathcal{P}_{db} - \left( g_d-\alpha_d \right)\mathcal{P}_{cb} \\ & \quad \; - \left( g_b-\alpha_b \right)\mathcal{P}_{[cd]} + \nabla_c \mathcal{P}_{db} - \nabla_d \mathcal{P}_{cb} \;, \end{split} \end{align}
\begin{align}
&\hat \cK_a^{\;\;b g r} =  \cK^{ \bar a}_{\ \bar b \bar g \bar r} \mathcal{G}_{a \bar a}^{\ \ \ b \bar b [g |\bar g| r] \bar r} \\
&\mathcal{G}_{\a \bar \a}^{\ \ \ \b \bar \b \g \bar \g \r \bar \r} = G_{\a \bar \a}G^{\b \bar \b}G^{\g \bar \g}G^{\r \bar \r} - 4\delta^\g_{\ \a}\delta^{\bar \g}_{\ \bar \a}G^{\b \bar \b}G^{\r \bar \r}
 + \delta^\g_{\ \a}\delta^{\bar \g}_{\ \bar \a}G^{\b \r}G^{\bar \b \bar \r}
\end{align}
\section{Field Equations in the Absence of  Matter Lagrangians}

The field equations for $\Pi\indices{^a _{bc}}$:
\begin{flalign}
&E_a{}^{mn} - \tfrac{1}{\rd+1} \delta_a{}^{(m}E_{b}{}^{n)b}  = 0\\
&E_a{}^{mn} = E_a{}^{nm} = \frac{1}{2\kappa_0 J_0 c}\breve\nabla_a\left( \sqrt{|g|}g^{mn} \right)-\breve\nabla_c\left( \sqrt{|g|} \hat \cK_{a}^{\,\,\, (m n) c} \right) \quad \nonumber\\
        &\qquad \qquad \qquad \quad+2 \l_0^2\breve\nabla_c\left( \sqrt{|g|}g_a \breve{\mathcal{K}}^{(mn)c}\right) -2  \l_0^2\sqrt{|g|}K^{(mn)c}\cD_{ca}
\end{flalign}

The field equations for $\mathcal{D}_{ab}$:
\begin{flalign}
&-\frac{1}{2\kappa_0 J_0 c} \sqrt{|g|} ({\rm{d}}-1) g^{pq} +\sqrt{|g|} \hat \cK_{c}^{\,\,\, (pq)c} + 2 \l_0^2 \breve\nabla_g( \sqrt{|g|} K^{(pq)g} ) - 2 \l_0^2 \sqrt{|g|} g_c \breve{\mathcal{K}}^{(pq)c} =0
\end{flalign}

The field equations for $g_{ab}$:
\begin{flalign}
\frac{1}{2}R_{(pq)} -&\frac{1}{2}Rg_{pq} =\kappa_0 \Theta^\text{S}_{pq}\\
\begin{split}  \Theta^\text{S}_{mn} =& -\tfrac{\rd-1}{2\kappa_0} \cP_{(mn)} + g_{mn} \left[2 J_0 c\l_0^2 (-\tfrac{1}{\rd+1} \nabla_a + \Delta_a) \cK^{a}{}_{bcd}K^{bcd}-\mathcal{L}_{\text{S}}\right] \cr
       &+ 2J_0c \lambda_0^2 (K_{mcd}K_n{}^{cd} + 2 K_{bcm}K^{bc}{}_n) + 2 J_0 c(8 \cK_{mb}\cK^b{}_n - 2 \cK \cK_{mn})\cr
       &+ 2 J_0 c (\cK_{mbcd}\cK_n{}^{bcd} -\cK^a{}_{mcd}\cK_{an}{}^{cd} - 2 \cK^{abc}{}_m \cK_{abcn})\cr
\mathcal{L}_{\text{S}}=&  - \frac{1}{2\kappa_0} ({\rm{d}}-1)\cP + c J_0 \lambda_0^{\ 2} K_{bcd}K^{bcd} 
-  c J_0  \left(\cK^{a}_{\ bcd}\cK_{a}^{\ bcd} - 4\cK_{ab}\cK^{ab} + \cK^2\right)
\end{split} \end{flalign}

\bibliographystyle{apsrev4-2}
\bibliography{Bibliography}
\end{document}